\documentclass[aps,prl,superscriptaddress, twocolumn, amsmath,amssymb,floatfix]{revtex4-2}
\usepackage{graphicx}
\usepackage{bm,siunitx, balance, color,soul}
\usepackage[colorlinks=True, allcolors=blue]{hyperref}
\usepackage[normalem]{ulem}
\usepackage{mathtools}

\begin{document}
    \title{The glass-forming ability of binary Lennard-Jones systems}

    \author{Yuan-Chao Hu}
    \affiliation{Department of Mechanical Engineering \& Materials Science, Yale University, New Haven, Connecticut 06520, USA}

    \author{Weiwei Jin}
    \affiliation{Department of Mechanical Engineering \& Materials Science, Yale University, New Haven, Connecticut 06520, USA}
   
    \author{Jan Schroers}
    \affiliation{Department of Mechanical Engineering \& Materials Science, Yale University, New Haven, Connecticut 06520, USA}

    \author{Mark D. Shattuck}
    \affiliation{Benjamin Levich Institute and Physics Department, The City College of New York, New York, New York 10031, USA.}

    \author{Corey S. O'Hern}
    \email[]{corey.ohern@yale.edu}
    \affiliation{Department of Mechanical Engineering \& Materials Science, Yale University, New Haven, Connecticut 06520, USA}
    \affiliation{Department of Physics, Yale University, New Haven, Connecticut 06520, USA.}
    \affiliation{Department of Applied Physics, Yale University, New Haven, Connecticut 06520, USA.}
    \date{\today}

\begin{abstract}
The glass-forming ability (GFA) of alloys, colloidal dispersions, and other particulate materials, as measured by the critical cooling rate $R_c$, can span more than ten orders of magnitude. Even after numerous previous studies, the physical features that control the GFA are still not well understood. For example, it is well-known that mixtures are better glass-formers than monodisperse systems and that particle size and cohesive energy differences among constituents improve the GFA, but it is not currently known how particle size differences couple to cohesive energy differences to determine the GFA. We perform molecular dynamics simulations to determine the GFA of equimolar, binary Lennard-Jones (LJ) mixtures versus the normalized cohesive energy difference $\epsilon_\_$ and mixing energy $\bar \epsilon_{AB}$ between particles A and B. We find several important results.  First, the $\log_{10} R_c$ contours in the $\bar \epsilon_{AB}$-$\epsilon_\_$ plane are ellipsoidal for all diameter ratios, and thus $R_c$ is determined by the Mahalanobis distance $d_M$ from a given point in the $\bar \epsilon_{AB}$-$\epsilon_\_$ plane to the center of the ellipsoidal contours. Second, LJ systems for which the larger particles have larger cohesive energy are generally better glass formers than those for which the larger particles have smaller cohesive energy.  Third, $d_M(\epsilon_\_,\bar \epsilon_{AB})$ is determined by the relative Voronoi volume difference between particles and local chemical order $S_{AB}$, which gives the average fraction of nearest-neighbor B particles surrounding an A particle and vice-versa. In particular, the shifted Mahalanobis distance $d_M - d^0_M$ versus the shifted chemical order $S_{AB}-S_{AB}^0$ collapses onto a hyperbolic master curve for all diameter ratios. These results identify design guidelines for improving the GFA of binary mixtures containing particles with different sizes, cohesive and mixing energies. 

\end{abstract}

\maketitle

\section*{I. introduction}

Nearly all materials can form amorphous solids, however, it is much 
more difficult for some materials to resist crystallization than
others.  Poor glass formers, such as pure metals, must be cooled
extremely rapidly (i.e. $ \geq 10^{14}\ \rm K/s$) to form amorphous structures~\cite{zhong_formation_2014}.  In
contrast, good glass formers, such as multi-component alloys, can 
form amorphous structures with cooling rates that are $12$ orders of magnitude slower~\cite{wang_bulk_2004,takeuchi_classification_2005,lu_new_2002,johnson_quantifying_2016}. 
Similarly, monodisperse colloidal dispersions can crystallize
in minutes, while specifically designed polydisperse colloidal systems
do not crystallize over week or year time scales~\cite{auer_suppression_2001,tanaka_bond_2012}.  
An important, open problem is determining the atomic- or particle-scale properties that control the glass-forming ability (GFA) of these materials.

Bulk metallic glasses have shown promise as structural materials and other applications since they possess large fracture toughness, high strength at elevated temperatures, and 
the ability to be processed like plastics~\cite{demetriou_damage-tolerant_2011,li_high-temperature_2019,schroers_processing_2010,ashby_metallic_2006}. However, a significant limitation to the widespread use of metallic glasses is that it is difficult to fabricate them as bulk samples~\cite{wang_bulk_2004,johnson_is_2015}.  
Numerous metallic glasses can only be formed as thin films and ribbons~\cite{ding_combinatorial_2014,li_how_2017,li_data-driven_2022}.  Further, bulk metallic glasses often contain precious metals and are expensive to produce~\cite{nishiyama_glass-forming_2002}. Thus, an important technological goal is to develop new alloys made from abundant elements that can form bulk metallic glasses. 

Numerous recent studies have shown that colloidal crystals can possess interesting optical applications such as photonic band gaps and wave guides~\cite{colvin_opals_2001, fudouzi_colloidal_2003,goerlitzer_bioinspired_2018}. Also, both disordered and ordered self-assembled colloidal structures give rise to vibrant structural coloration in many biological systems~\cite{goerlitzer_bioinspired_2018}, such as insects and birds. Thus, understanding the physical parameters that control crystallization versus glass-formation will aid in the development of colloidal assemblies with targeted structural and optical properties.      

A common technique for studying glass formation is to begin in the liquid state and cool 
the liquid phase
at different rates to determine its susceptibility to crystallization~\cite{kai_zhang_computational_2013}. A first approach for modeling the liquid states of colloids, alloys, and other
glass-forming materials is to describe them as ``simple liquids".  A
simple liquid is a collection of $N$ spherical atoms or particles that interact via classical, pairwise or multi-body potentials. Typical pairwise interatomic potentials for simple liquids, such as the Lennard-Jones (LJ) and Morse
potentials, include parameters for the diameter of the particles
$\sigma$ and the depth $\epsilon$ of the attractive interactions
between particles. Note that the properties of transition metals have been described previously using Lennard-Jones interactions ~\cite{hu_glass_2020}.  In general, it is well-known that mixtures of
different types of particles are better glass formers than monodisperse systems~\cite{lu_new_2002,li_how_2017}.  For example, binary systems are typically better glass-formers than systems with a single particle type, ternary systems are typically better glass-formers than binary
systems, and so on.  For binary LJ systems with the same 
number of small and large particles, particles with size ratios of $0.7$-$0.8$ form 
amorphous structures without de-mixing over a wide range of cooling rates~\cite{zhang_connection_2014,zhang_beyond_2015}. For non-equimolar, multi-component LJ systems, studies suggest that the number fraction of the different-sized particles should be determined such that the total particle volume of each species is the same to ensure amorphous structures that do not de-mix and crystallize during cooling~\cite{zhang_connection_2014, zhang_beyond_2015}. 

In general, differences in particle-scale properties improve the glass-forming ability of mixtures~\cite{gfa_2019,hu_glass_2020}. In previous studies of LJ systems composed of the same-sized particles, we identified two dimensionless energetic parameters, the normalized cohesive energy difference, $\epsilon_\_ = (\epsilon_{BB}-\epsilon_{AA})/(\epsilon_{BB}+\epsilon_{AA})$ and mixing energy, $\bar \epsilon_{AB} = 2\epsilon_{AB}/(\epsilon_{BB}+\epsilon_{AA})$, that control the glass-forming ability~\cite{gfa_2019}. 
However, little is known about the glass-forming ability of mixtures with different particle sizes {\it and} different attractive interactions. For example, suppose we have a binary mixture of A and B particles with diameters $\sigma_{AA}$ and $\sigma_{BB}$, cohesive energies $\epsilon_{AA}$ and $\epsilon_{BB}$, mixing energy $\epsilon_{AB}$, and number fraction of B particles $f_B$. What combination of these parameters gives rise to the best glass-forming ability?  Specifically, is the system with a larger diameter $\sigma_{AA} > \sigma_{BB}$ and larger cohesive energy $\epsilon_{AA} > \epsilon_{BB}$ a better glass-former than the system with $\sigma_{AA} > \sigma_{BB}$ and smaller cohesive energy $\epsilon_{AA} < \epsilon_{BB}$?  Also, can we improve the glass-forming ability by tuning $\epsilon_{AB}$? 

To address these questions, we carry out extensive molecular dynamics simulations to determine the critical cooling rate $R_c$ (minimum rate above which crystallization does not occur) of binary LJ mixtures over a wide range of $\epsilon_\_$ and $\bar \epsilon_{AB}$. We focus on equimolar mixtures with $f_B=0.5$ and several particle size ratios. We find four key results. First, the critical cooling rate contours in the $\epsilon_\_$ and $\bar \epsilon_{AB}$ plane are ellipsoidal for all diameter ratios studied, and thus $R_c$ can be determined by the Mahalanobis distance $d_M$ from a given point in the $\bar \epsilon_{AB}$-$\epsilon_\_$ parameter space to the center of the ellipsoidal contours. Second, LJ systems for which the larger particles have larger cohesive energy (i.e. $\epsilon_{BB}/\epsilon_{AA} < 1$ and  $\sigma_{BB}/\sigma_{AA} < 1$) are generally better glass formers than those for which the larger particles have smaller cohesive energy (i.e. $\epsilon_{BB}/\epsilon_{AA} < 1$ and $\sigma_{BB}/\sigma_{AA} > 1$). We show that LJ systems for which the larger particles have larger cohesive energy possess inherent structures with lower potential energy and higher energy barriers compared to those for the opposite case. Third, the Mahalanobis distance $d_M$, and thus the glass-forming ability, for a given point in the $\epsilon_\_$ and $\bar \epsilon_{AB}$ plane is determined by the relative Voronoi volume difference between the A and B particles and the local chemical order $S_{AB}$~\cite{chemorder1960,hu_physical_2020}, which gives the average fraction of nearest-neighbor B particles surrounding an A particle and vice-versa. In particular, the shifted Mahalanobis distance $d_M - d^0_M$ versus the shifted chemical order $S_{AB}-S_{AB}^0$ collapses onto a hyperbolic master curve for all diameter ratios studied. Fourth, we show that the best LJ glass-formers display bond-shortening behavior, where the separation between weakly interacting particles is smaller than the value given by the minimum in the pair potential. These results identify several important design guidelines for improving the glass-forming ability of binary mixtures containing particles with different sizes, cohesive and mixing energies. 

The reminder of the article is organized into three sections. In Sec. II, we describe the computational methods including the interaction potential, molecular dynamics simulations, structural characterization, and measurements of the critical cooling rate, local chemical order, and Voronoi volumes. In Sec. III, we describe the main results of the work and discuss their importance and implications. In Sec. IV, we provide several promising future research directions, including computational studies of the glass-forming ability of binary mixtures over a range of compositions $f_B$ and generalizations of the work to ternary and quaternary systems.  The Appendix describes differences in the potential energy of the inherent structures and glass-forming ability of binary LJ systems with $\sigma_{BB}/\sigma_{AA} < 1$ and $\epsilon_{BB}/\epsilon_{AA} < 1$ versus those with $\sigma_{BB}/\sigma_{AA} > 1$ and $\epsilon_{BB}/\epsilon_{AA} < 1$.

\section*{II. Methods}

In this section, we describe the computational methods that we employ to investigate the glass-forming ability of binary LJ mixtures, including the molecular dynamics simulations and the cooling protocol, characterization of local structure, measurement of the critical cooling rate, and analyses of local chemical order. 

\subsection{A. Molecular Dynamics Simulations}

\begin{figure}[b!]
\centering
\includegraphics[width = \linewidth]{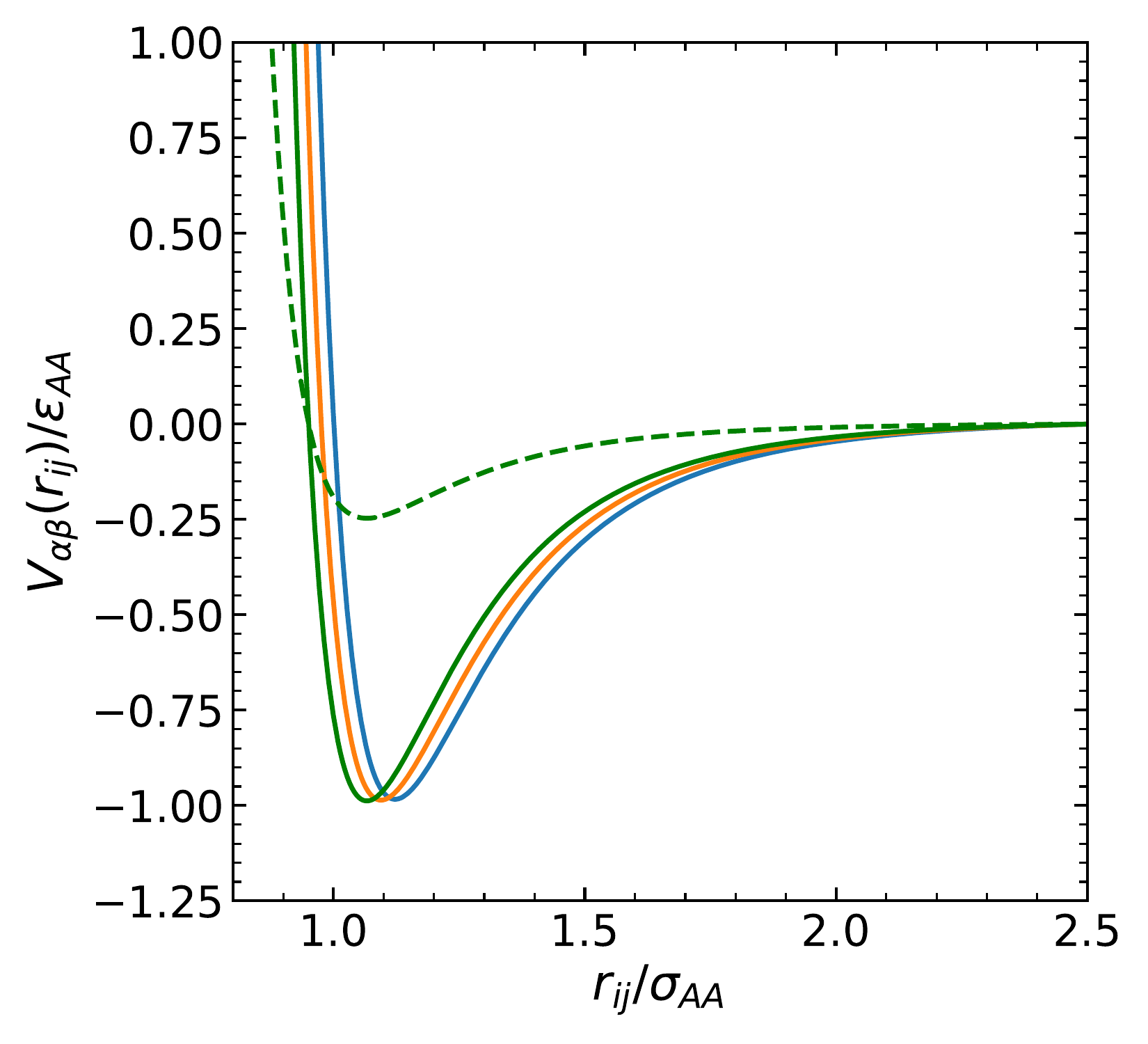}
\caption{
The binary Lennard-Jones pair potential $V_{\alpha \beta}(r_{ij})/\epsilon_{AA}$ for diameter ratio $\sigma_{BB}/\sigma_{AA}=0.95$ and two sets of energetic parameters:
$\epsilon_{BB}/\epsilon_{AA}=\epsilon_{AB}/\epsilon_{AA}=1.0$ (solid lines) and  
$\epsilon_{BB}/\epsilon_{AA}=0.25$ and $\epsilon_{AB}/\epsilon_{AA}=1.0$ (dashed lines). The $V_{AA}$, $V_{AB}$, and $V_{BB}$ pair potentials are represented by blue, orange, and green lines, respectively. $V_{AA}$ and $V_{AB}$ are the same for both sets of energetic parameters.}
\label{fig0}
\end{figure}

We focus on binary systems composed of equal number fractions of A and B particles ($f_A=f_B=0.5$) with equal mass $m$. We assume that the particles interact through the pairwise Lennard-Jones potential:
\begin{equation}
    V_{\alpha \beta} (r_{ij}) = 4 \epsilon_{\alpha \beta} \left[ \left( \frac{\sigma_{\alpha \beta}}{r_{ij}}\right)^{12} - \left( \frac{\sigma_{\alpha \beta}}{r_{ij}} \right)^{6} \right],
\end{equation}
where $\alpha, \beta$ indicate which of the particles (A or B) are interacting and $r_{ij}$ is the separation between particles $i$ and $j$.  We consider systems confined to cubic boxes with periodic boundary conditions in the $x$-, $y$-, and $z$-directions and $N=2000$ is the total number of particles. (In previous studies of the glass-forming ability of binary Lennard-Jones systems, we showed that the finite-size effects for the critical cooling rate for $N>10^3$ are weak~\cite{gfa_2019}.) We consider an additive mixing model for the particle diameters 
$\sigma_{AB} = (\sigma_{AA} + \sigma_{BB})/2$,
but we vary $\epsilon_{AB}$ independently relative to $\epsilon_{BB}$ and $\epsilon_{AA}$. We choose length and energy units such that $\sigma_{AA}=\sigma$ and $\epsilon_{AA}=\epsilon > \epsilon_{BB}$~\cite{gfa_2019}.  We consider systems with $\sigma_{BB}/\sigma_{AA}=1.05$, $0.99$, $0.97$, and $0.95$, which allows us to crystallize the systems over a wide range of the energetic parameter space. The Lennard-Jones potential is truncated and shifted at $2.5\sigma_{\alpha \beta}$. The Lennard-Jones pair potential $V_{\alpha \beta}(r_{ij})$ for diameter ratio $\sigma_{BB}/\sigma_{AA}=0.95$ and two sets of energetic parameters is shown in Fig.~\ref{fig0}. The pressure, temperature, and time scales are reported in units of $\epsilon/\sigma^3$, $\epsilon/k_b$, and $\sqrt{m \sigma^2 / \epsilon}$, where $k_b$ is the Boltzmann constant. 

To investigate the glass-forming ability of binary Lennard-Jones systems, we first equilibrate them at a high temperature $T=5.0$ above the glass transition temperature $T_g$ and then quench them to low temperature $T=0.1$ below $T_g$ over a range of linear cooling rates $R$. The simulations are carried out in the isothermal-isobaric (NPT) ensemble using the Nos\'{e}-Hoover thermostat and barostat with a pressure $p=10$, which enables the system to avoid cavitation over the full range of parameters. The equations of motion are integrated using a modified velocity-verlet algorithm with time step $\Delta t=2 \times 10^{-3}$ and the time constants for the thermostat and barostat are set to $10^2 \Delta t$ and $10^3 \Delta t$, respectively.

\subsection{B. Characterization of Local Structure}

To determine the critical cooling rate $R_c$, we analyze the local structural order of the low temperature solids by quantifying the bond orientational order for each particle~\cite{steinhardt_bond-orientational_1983,tanaka_bond_2012}. The nearest neighbors of each particle are obtained by performing Voronoi tessellation~\cite{rycroft_analysis_2006}.  We calculate the bond orientational order parameter $q_{6m}(i)$ for each particle $i$:
\begin{equation}
q_{6m}(i) = \sum_{j=1}^{N_i} \frac{A_j}{A^i_{\rm tot}} Y_{6m}(\theta(\mathbf{r}_{ij}), \phi(\mathbf{r}_{ij})),
\end{equation}
where $N_i$ is the number of nearest Voronoi neighbors of particle $i$, $Y_{6m}(\theta({\mathbf r}_{ij},\phi({\mathbf  r}_{ij}))$ is the spherical harmonic function of degree $6$ and order $m$, and $\theta$ and $\phi$ are the polar and azimuthal angles. The contribution from the spherical harmonics of each neighbor $j$ of particle $i$ is weighted by the fraction $A_j/A^i_{\rm tot}$ of the area of the Voronoi face separating the two particles to the total area of all faces $A^i_{\rm tot}$ of the polyhedron surrounding particle $i$. 
We determine the number of crystal-like atoms by calculating the correlations in the bond orientational order parameter:
\begin{equation}
s_6(i,j) = \frac{\sum\limits_{m=-6}^6 q_{6m}(i) q_{6m}^\ast(j)}{\sqrt{\sum\limits_{m=-6}^6 {\vert q_{6m}(i) \vert}^2 
\sum\limits_{m=-6}^6 {\vert q_{6m}(j) \vert}^2}}, 
\end{equation}
where $q_{6m}^\ast(j)$ is the complex conjugate of $q_{6m}(j)$.
If $s_6(i,j) > 0.7$, we treat the bond as crystal-like~\cite{rein_ten_wolde_numerical_1996}. If the total number of crystal-like bonds for a given particle is larger than $10$, the particle is considered to be in a crystalline environment. The sensitivity of the thresholds for $s_6(i,j)$ and the number of crystal-like bonds have been studied previously~\cite{rein_ten_wolde_numerical_1996,russo_microscopic_2012}. A benefit of using $s_6(i,j)$ is that it does not require specifying the local symmetry of each crystalline phase to determine the crystalline particles~\cite{hu_glass_2020}. For each set of size ratios and energetic parameters, we calculate the fraction of crystalline particles $f_{\rm xtal}$ as a function of the cooling rate $R$. 

\begin{figure}[t!]
\centering
\includegraphics[width = \linewidth]{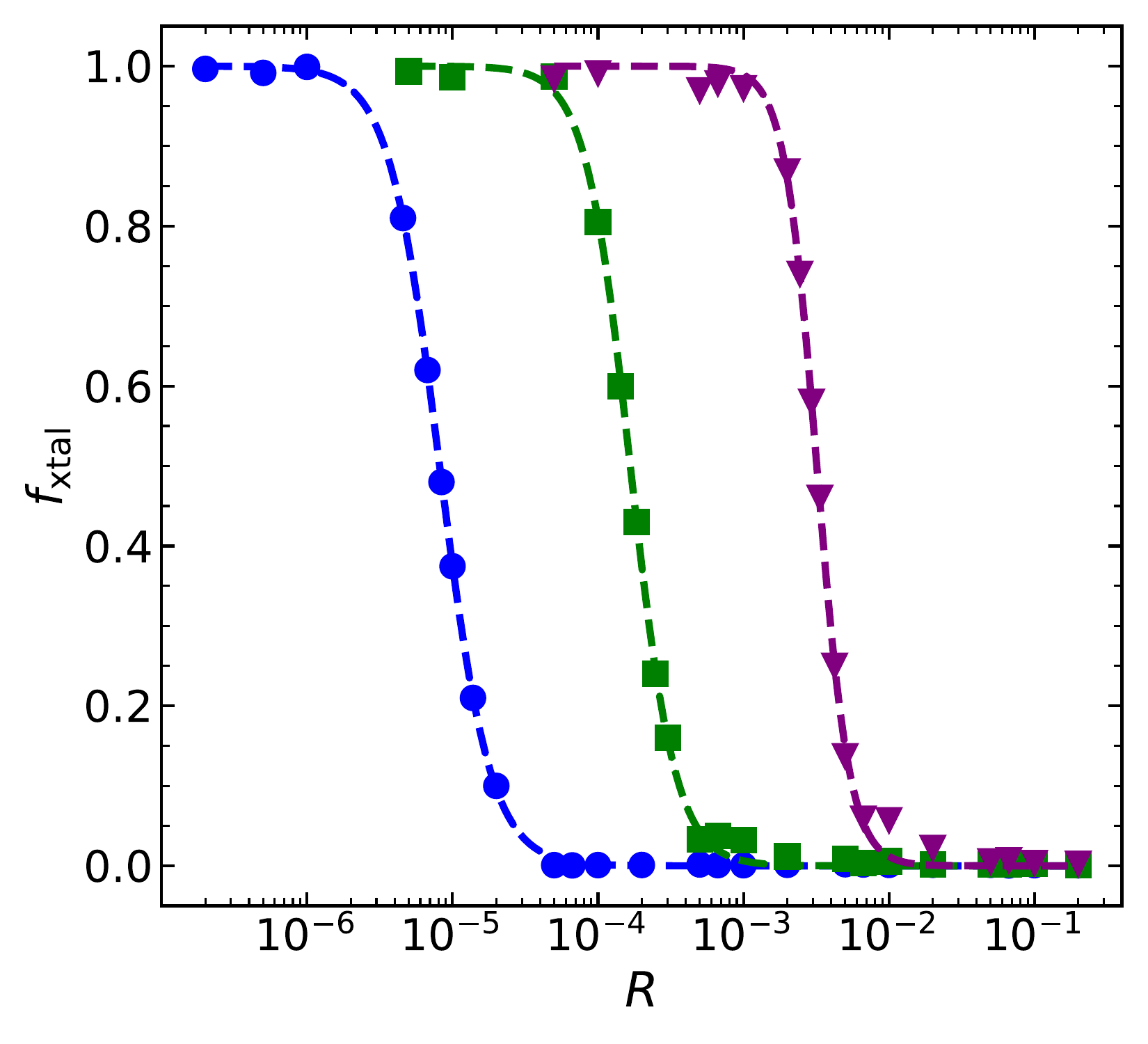}
\caption{The fraction of crystalline particles $f_{\rm xtal}$ in the low-temperature solid plotted as a function of cooling rate $R$ for Lennard-Jones binary mixtures with diameter ratio $\sigma_{BB}/\sigma_{AA}=0.95$.  The data was averaged over $30$ independent trials cooling from $T= 5$ above the glass transition temperature $T_g$ to $T=0.1$ below $T_g$. The dashed lines are best fits to Eq.~\ref{sigmoid}, which allows us to determine the critical cooling rate $R_c$ at which $f_{\rm xtal} = 0.5$. We show data for three sets of parameters:  ${\overline \epsilon}_{AB}=1.23$ and $\epsilon_\_=-0.538$ (circles); ${\overline \epsilon}_{AB}=1.33$ and $\epsilon_\_=-0.333$ (squares), and ${\overline \epsilon}_{AB}=1.0$ and $\epsilon_\_=0$ (triangles).
 }
\label{fig1}
\end{figure}

\begin{figure*}[t!]
\centering
 \includegraphics[width = \textwidth]{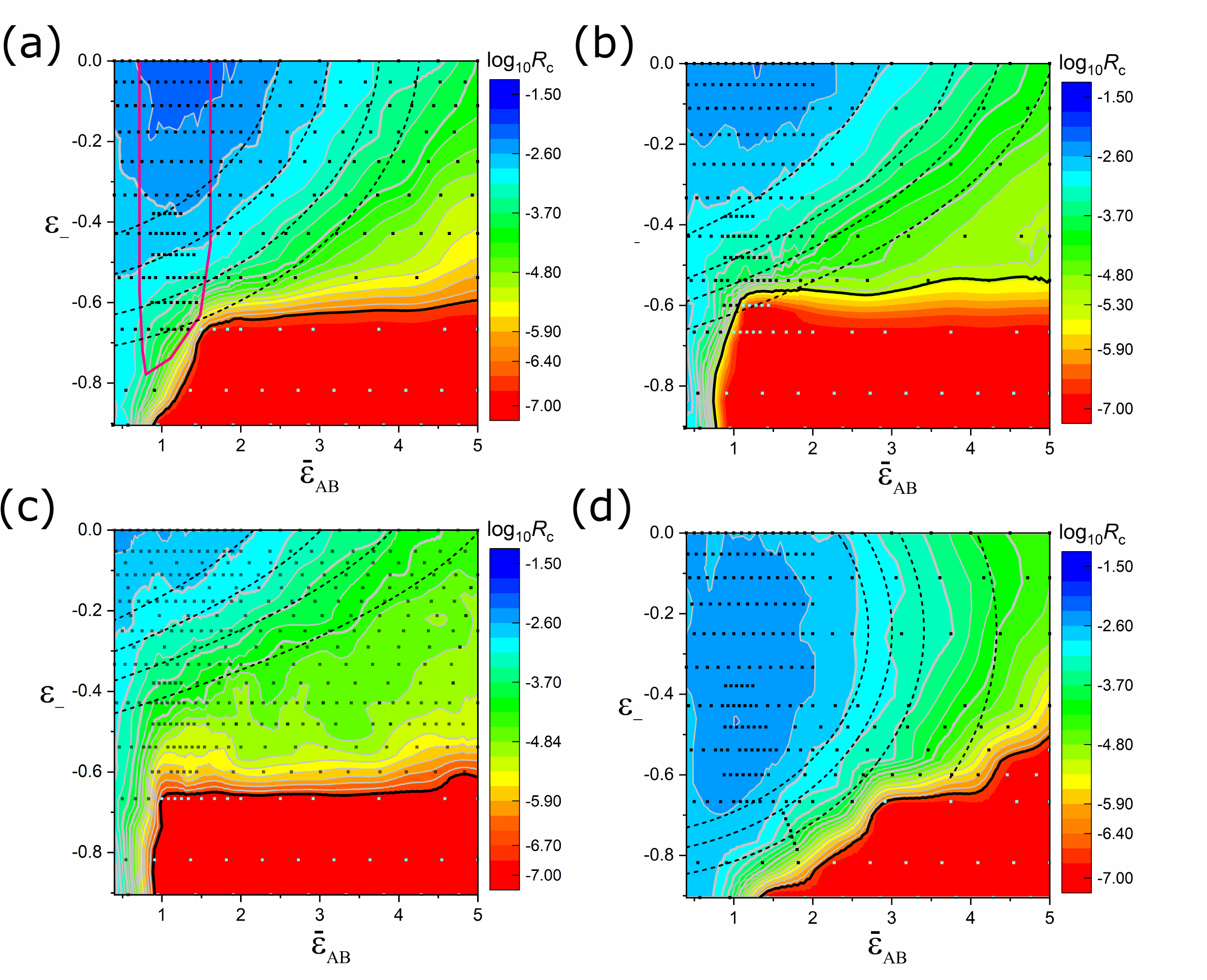}
 \caption{
Contour plots of the critical cooling rate $R_c$ as a function of the normalized cohesive energy difference $\epsilon_\_$ and mixing energy ${\overline \epsilon}_{AB}$ for several diameter ratios: (a) $\sigma_{BB}/\sigma_{AA}=0.99$, (b) $0.97$, (c) $0.95$, and (d) $1.05$.  The black squares indicate the values of $\epsilon_\_$ and ${\overline \epsilon}_{AB}$ that were sampled using the MD simulations and these systems can be crystallized using the cooling rates we considered. $R_c$ decreases by more than five orders of magnitude as the color changes from blue to red. The parameter space enclosed by the solid magenta line in (a) indicates the region where binary alloys occur. The light gray solid lines indicate several constant-$\log_{10} R_c$ contours. The solid black line marks the lowest cooling rate that we studied. The black dashed lines are best fits of Eq.~\ref{ellipse} to a few selected $\log_{10} R_c$ contour lines (bold gray lines).
The light cyan squares indicate values of $\epsilon_\_$ and ${\overline \epsilon}_{AB}$ that were sampled, but we were not able to crystallize these systems.}
\label{fig2}
\end{figure*}

\subsection{C. Measurement of the Critical Cooling Rate}

In general, $f_{\rm xtal}$ versus the logarithm of the cooling rate, $\log_{10} R$, is a sigmoidal function, where $f_{\rm xtal} \approx 1$ as $R \rightarrow 0$ and $f_{\rm xtal} \approx 0$ as $R \rightarrow \infty$. To measure the critical cooling rate $R_c$ at which $f_{\rm xtal} = 0.5$, we assume that 
\begin{equation}
\label{sigmoid}
    f_{\rm xtal} = \frac{1}{2} \left( 1 - \tanh \left[ \log_{10} (R/R_c)^{-\kappa} \right] \right),
\end{equation}
where $0 < \kappa < 1$ is the stretching exponent~\cite{gfa_2019,hu_glass_2020}. We show examples of $f_{\rm xtal}$ versus $R$ for three sets of diameter ratios and energy parameters in Fig.~\ref{fig1}. $R_c$ varies by more than a factor of $100$ over this range of parameters. 

\subsection{D. Analysis of Chemical Order}

It is well-known that the local composition of dense liquids and glasses can deviate strongly from the nominal, or globally averaged, composition (i.e. $f_A=f_B=0.5$ in the current study). Deviations from the nominal composition are frequently termed local ``chemical order"~\cite{chemorder1960}. To quantify local chemical order, we measure the average fraction of particles ${\overline S}_{\alpha \beta}$ of type $\beta$ that are Voronoi neighbors of particles of type $\alpha$. To make this quantity symmetric, we define 
$S_{AB} = ({\overline S}_{AB} + {\overline S}_{BA})/2$. 
Note that $S_{\alpha \beta}$ is coupled to local packing, since the number of A-type nearest neighbors surrounding a B particle and vice versa are affected by the diameter ratio and local density, as well as the energetic parameters.

\section*{III. Results}

In this section, we describe the results from the molecular dynamics simulations of thermally quenched Lennard-Jones binary mixtures as a function of the diameter ratio and energetic parameters. As discussed in Sec. II.C, we calculate the fraction of crystalline atoms $f_{\rm xtal}$ versus the cooling rate $R$, and thus we can determine the critical cooling rate $R_c$, for each diameter ratio and combination of the energetic parameters, $\epsilon_{BB}/\epsilon_{AA}$ and $\epsilon_{AB}/\epsilon_{AA}$. In previous work~\cite{gfa_2019}, we showed that plotting $R_c$ versus the normalized interaction energy $\bar \epsilon_{AB} = 2\epsilon_{AB}/(\epsilon_{BB}+\epsilon_{AA})$ and cohesive energy difference $\epsilon_\_=(\epsilon_{BB}-\epsilon_{AA})/(\epsilon_{BB}+\epsilon_{AA})$ can provide improved collapse of the data. We show contours of $\log_{10} R_c$ versus $\bar \epsilon_{AB}$ and $\epsilon_\_$ for four different diameter ratios in Fig.~\ref{fig2}.  Note that we sample a much larger energetic parameter space than that occupied by binary alloys (region bounded by the magenta solid line) in Fig.~\ref{fig2} (a). 

Figure~\ref{fig2} illustrates several key points. First, $R_c$ decreases with decreasing $\epsilon_\_$ (i.e. as $\epsilon_\_$ becomes more negative) and increasing $\bar \epsilon_{AB}$. Second, the glass-forming ability dramatically improves with decreasing diameter ratio. We find that the blue region with large values of $R_c$ for the system with $\sigma_{BB}/\sigma_{AA} =0.99$ (Fig.~\ref{fig2} (a)) takes up a much larger region of the $\bar \epsilon_{AB}$-$\epsilon_\_$ parameter space than that for the system with $\sigma_{BB}/\sigma_{AA} =0.95$ (Fig.~\ref{fig2} (c)). Third, there is a significant difference in the $\log_{10} R_c$ contours between the two similar systems: $\sigma_{BB}/\sigma_{AA}=0.95$ and $\epsilon_{BB}<\epsilon_{AA}$ (Fig.~\ref{fig2} (c)) and  $\sigma_{BB}/\sigma_{AA}=1.05$ and $\epsilon_{BB}<\epsilon_{AA}$ (Fig.~\ref{fig2} (d)). In the former case, the larger particle possesses the larger cohesive energy and in the latter case, the smaller particle possesses the larger cohesive energy.  In general, for each point in the $\bar \epsilon_{AB}$-$\epsilon_\_$ parameter space, the glass-forming ability is improved if the larger particle possesses the larger cohesive energy. (In the Appendix, we show that these better glass-forming systems possess deeper energy minima and larger energy barriers.) This finding is consistent with four out of six binary bulk metallic glasses, such as Cu-Zr and Ni-Nb, and $31$ out of $47$ ribbon-forming metallic glasses~\cite{li_how_2017}. Choosing elements such that the larger atom possesses larger cohesive energy is a novel design principle for the development of bulk metallic glasses. 

\begin{table}[b!]
\caption{\label{tab:table1}%
The position of the ellipse center $(\bar \epsilon_{AB}^0, \epsilon^0_\_)$, major and minor axes $a$ and $b$, as well as the reference cooling rate $\log_{10} R_0$ from best fits of Eq.~\ref{ellipse} to the ellipsoidal contours in Fig. 3 for all diameter ratios $\sigma_{BB}/\sigma_{AA}$.} 
\begin{ruledtabular}
\begin{tabular}{cccccccc}
$\sigma_{BB}/\sigma_{AA}$ & $(\bar \epsilon_{AB}^0, \epsilon^0_\_)$ & $a$ & $b$ & $\log_{10} R_0$ \\
\hline
$1.05$ & (-0.538, -0.243) & 3.182 & 0.501 & -1.962 \\
$0.99$ & (-1.065, 0.046) & 3.701 & 0.544 & -1.724 \\
$0.97$ & (-7.328, 0.103) & 6.405 & 0.501 & -0.434 \\
$0.95$ & (-30.598, 0.218) & 13.696 & 0.496 & 2.912\\
\end{tabular}
\end{ruledtabular}
\end{table}

Another feature of the $\log_{10} R_c$ contours in Fig.~\ref{fig2} is that they have ellipsoidal shapes in the $\bar \epsilon_{AB}$-$\epsilon_\_$ parameter space with the major axis oriented in the $\bar \epsilon_{AB}$ direction. Therefore, we used the following ellipsoidal form to describe all of the data in Fig.~\ref{fig2}:
\begin{equation}
\begin{split}
\log_{10} \left( \frac{R_c}{R_0}\right) 
&= -\left[
\frac{(\bar \epsilon_{AB} - \bar \epsilon_{AB}^0)^2}{a^2} + \frac{(\epsilon_\_ - \epsilon_\_^0)^2}{b^2}\right] \\
&=-d_M^2(\bar \epsilon_{AB},\epsilon_\_),
\end{split}
\label{ellipse}
\end{equation}
where $R_0$ is a reference cooling rate, $(\bar \epsilon_{AB}^0, \epsilon^0_\_)$ gives the ellipse center, and $a$ and $b$ are the major and minor axes of the ellipse, respectively. $d_M$ is the Mahalanobis distance, which gives the separation between a point in the $\bar \epsilon_{AB}$-$\epsilon_\_$ plane to the center of the ellipse. Several examples of the ellipsoidal fits are shown as the dashed lines in Fig.~\ref{fig2}. The parameters of the ellipsoidal contours are provided in Table~\ref{tab:table1}. We find that the lengths of the minor axes remain nearly constant as we tune the diameter ratio, whereas the length of the major axis increases by a factor of $\sim 4$ as we decrease the diameter ratio. In addition, the center of the ellipse shifts to large negative values of $\bar \epsilon_{AB}$ with decreasing diameter ratio, which contributes to the strong increase in the glass forming ability. 

\begin{figure}[t!]
\centering
\includegraphics[width=\linewidth]{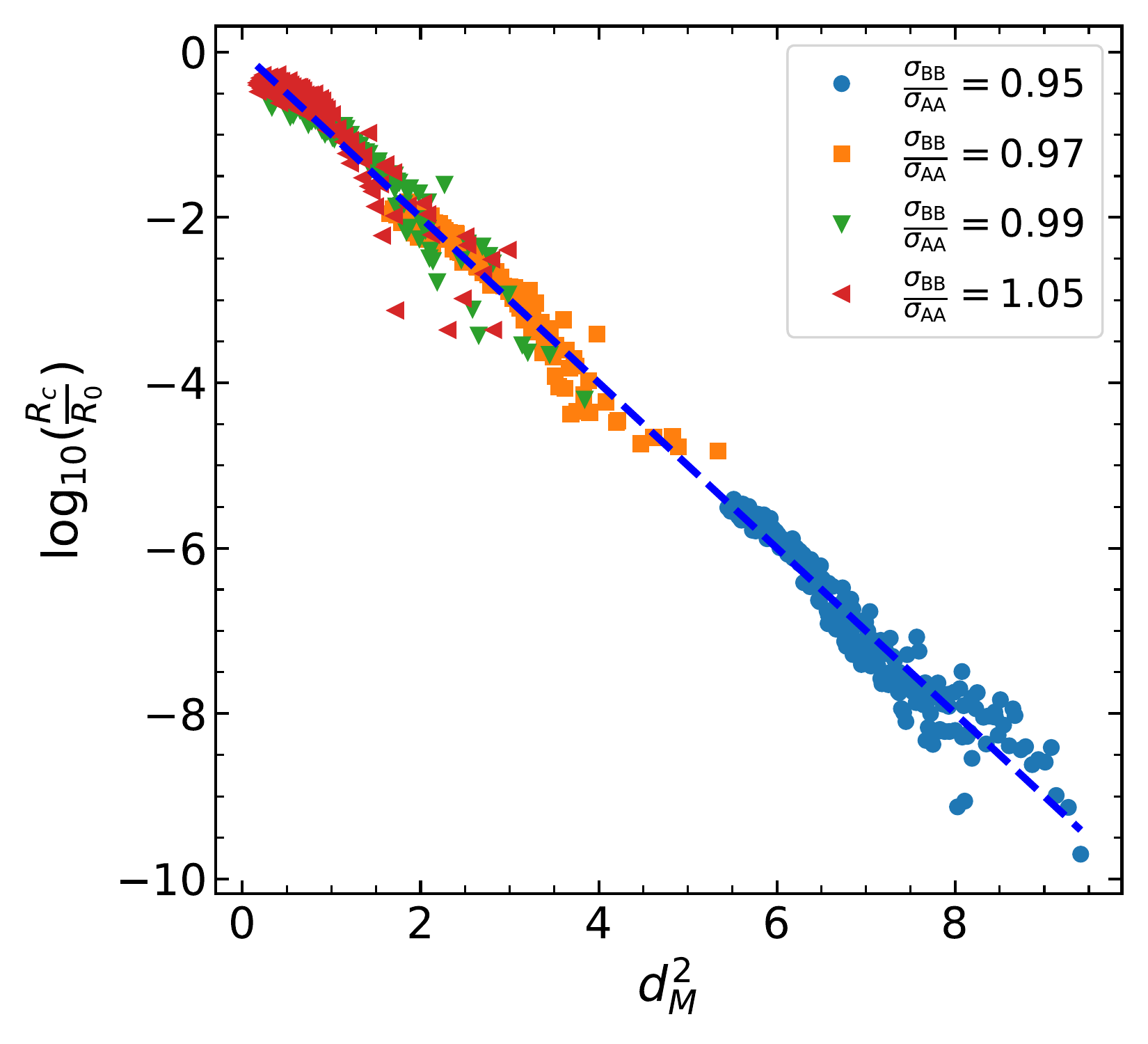}
\caption{
The critical cooling rate $R_c$ obtained from the molecular dynamics simulations (normalized by $R_0$) plotted versus the square of the Mahalanobis distance $d^2_M$, where $d_M$ is the separation between a given point in the $\bar \epsilon_{AB}$-$\epsilon_\_$ plane to the ellipse center for each diameter ratio. The dashed line gives $\log_{10} R_c/R_0 = -d^2_M$.}
\label{fig3}
\end{figure}

The quality of the fits of the $\log_{10} R_c$ contours to Eq.~\ref{ellipse} is assessed in Fig.~\ref{fig3}, where we plot $\log_{10} (R_c/R_0)$ versus the square of the Mahalanobis distance, $d_M^2$. We show that as the diameter ratio decreases, $d_M$ increases and $R_c/R_0$ decreases. The R-squared value for the fit of the data in Fig.~\ref{fig3} to Eq.~\ref{ellipse} is $\sim 0.99$, and thus the ellipsoidal approximation in Eq.~\ref{ellipse} is a high-quality description of the $\log_{10} R_c$ contours. Thus, the glass-forming ability of binary Lennard-Jones systems can be characterized by $d_M(\bar \epsilon_{AB},\epsilon_\_)$ and a diameter ratio-dependent offset $R_0$.

To understand the particle-scale features that determine $d_M$ (and the glass-forming ability, $R_c/R_0$) for each binary Lennard-Jones system, we characterize the structural properties of glassy solids obtained after rapid cooling (i.e. using a cooling rate $R=10^{-2}$, which is much larger than $R_c$ for all of the systems studied).  In previous work on monoatomic Lennard-Jones systems~\cite{gfa_2019}, we found that the chemical order obtained from rapidly cooled systems at each point in the $\bar \epsilon_{AB}$-$\epsilon_\_$ plane can provide significant insight into the glass-forming ability.  In particular, we showed that deviations in the local composition from the nominal value are correlated with enhanced glass-forming ability. In the current work, we will correlate both the chemical order $S_{AB}$ and relative Voronoi volume difference between the two particle types of the rapidly cooled glassy solids with the glass-forming ability at each point in the $\bar \epsilon_{AB}$-$\epsilon_\_$ plane.  

\begin{figure*}[t!]
\centering
\includegraphics[width = \textwidth]{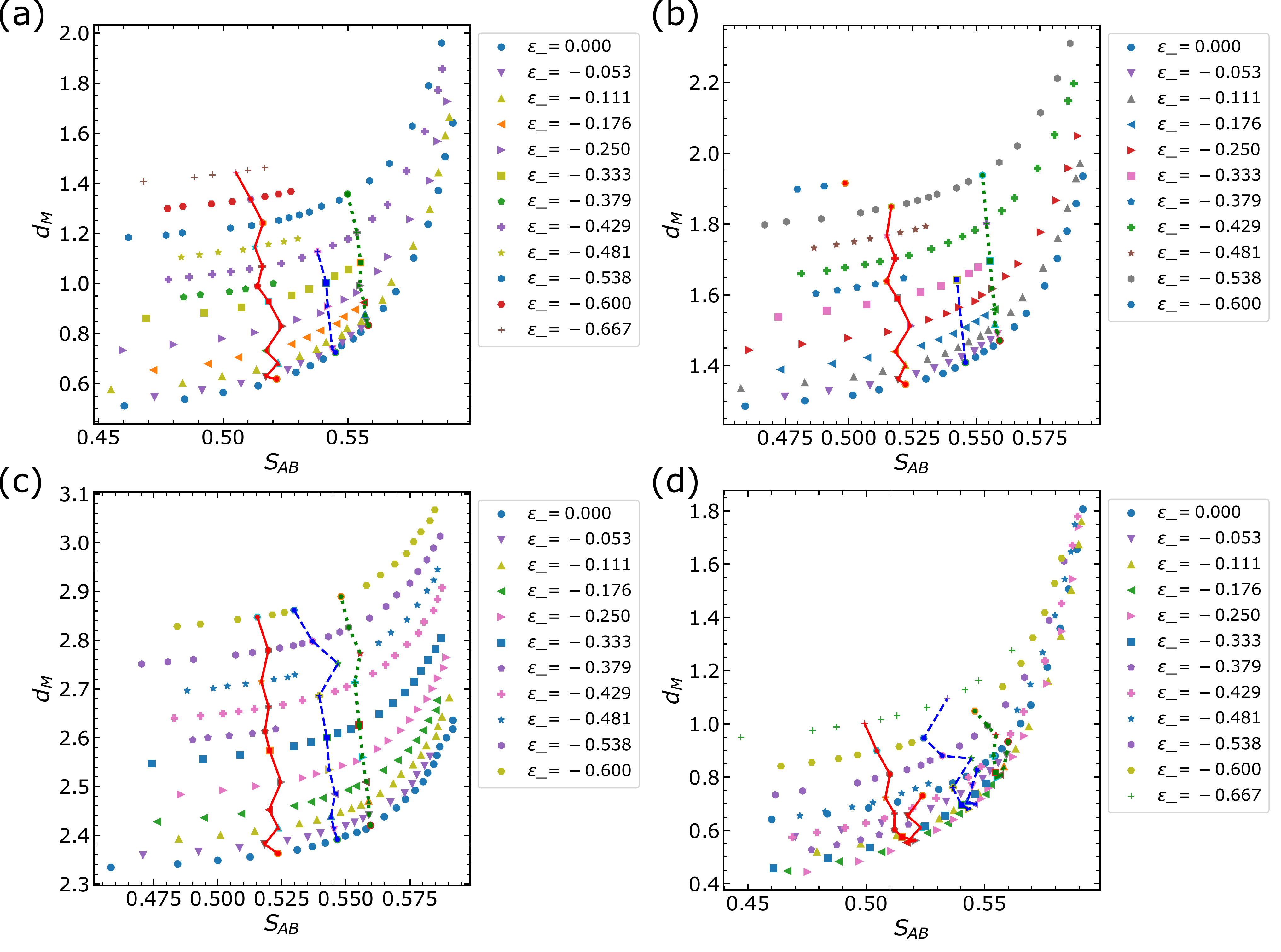}
\caption{
The Mahalanobis distance $d_M$ plotted versus the local chemical order $S_{AB}$ for rapidly cooled ($R=10^{-2})$ binary Lennard-Jones systems with diameter ratios (a) $\sigma_{BB}/\sigma_{AA}=0.99$, (b) $0.97$, (c) $0.95$, and (d) $1.05$. The symbols in each panel indicate systems at different values of $\epsilon_\_$. The red solid lines, blue dashed lines, and green dotted lines in each panel indicate samples at fixed $\bar \epsilon_{AB} \approx 1.2$, $1.6$, and $2.0$, respectively. }
\label{fig4}
\end{figure*}

\begin{figure}[b!]
\centering
 \includegraphics[width = \linewidth]{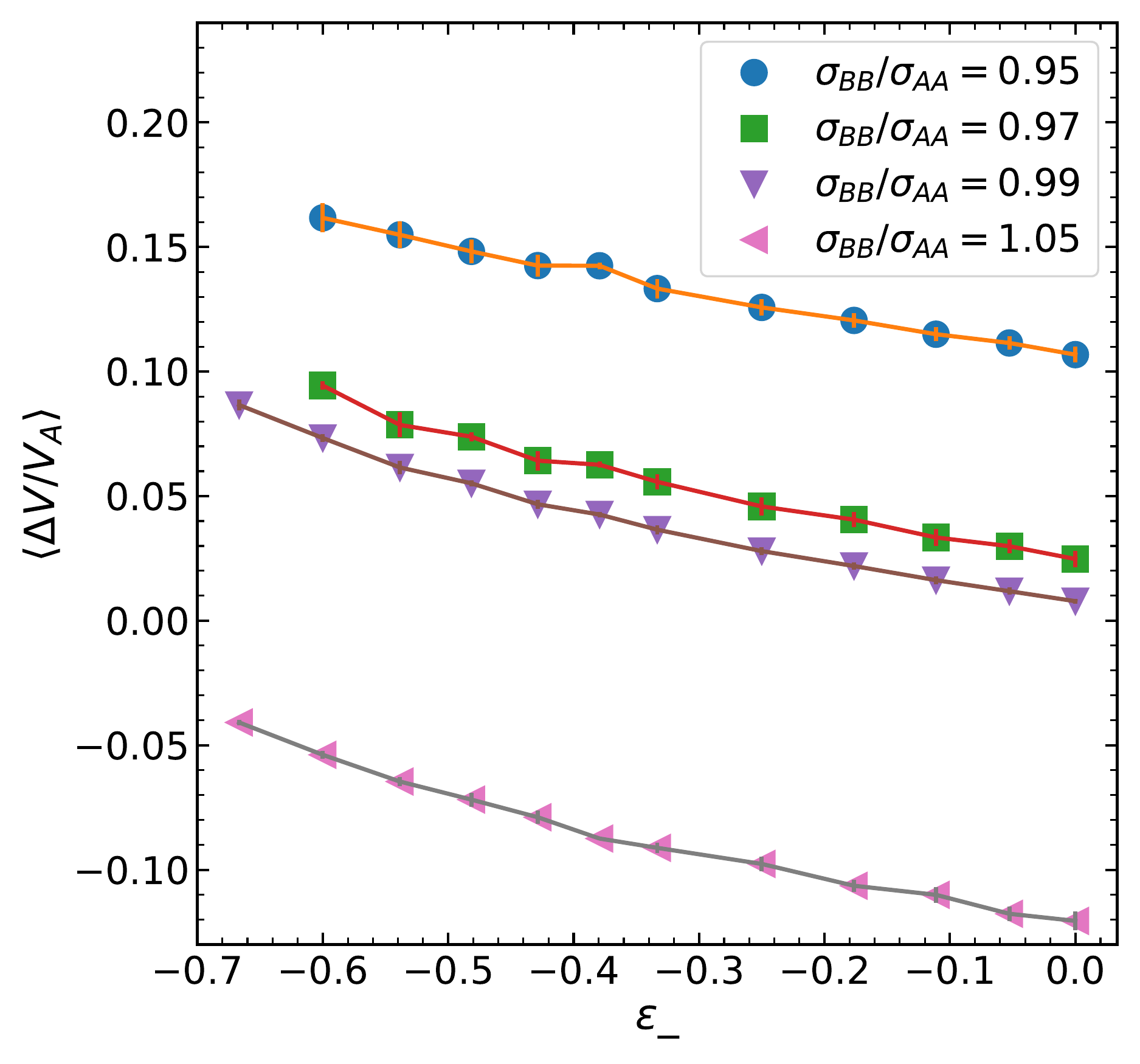}
 \caption{
The relative difference between the average Voronoi volumes of particle types A and B, $\langle \Delta V/V_A \rangle = \langle (V_A - V_B)/V_A \rangle$ plotted versus $\epsilon_\_$ for all four diameter ratios. Each data point is obtained by averaging $\langle \Delta V/V_A \rangle$ over all values of $\bar \epsilon_{AB}$ at fixed $\bar \epsilon_\_$. The error bars represent the standard deviation at each value of $\epsilon_\_$.}
\label{fig5}
\end{figure}

In Fig.~\ref{fig4}, we show $d_M$ versus the local chemical order $S_{AB}$ for the LJ systems with diameter ratios $\sigma_{BB}/\sigma_{AA} = 0.99$, $0.97$, and $0.95$, where the larger particles have larger cohesive energies, and for $\sigma_{BB}/\sigma_{AA} = 1.05$, where the larger particles have smaller cohesive energies. We organize the data into groups at fixed values of $\epsilon_\_$ and varying $\bar \epsilon_{AB}$.  For each system, as $\bar \epsilon_{AB}$ increases, $S_{AB}$ and $d_M$ increase. Thus, increases in the mixing energy $\bar \epsilon_{AB}$ give rise to deviations in the local composition from the nominal value, which enhance the glass-forming ability. In each case, as $S_{AB}$ increases above $0.5$, there is a rapid increase in $d_M$. Figure~\ref{fig4} also shows that there is an $S_{AB}$-independent offset to $d_M$ that increases as $\epsilon_\_$ becomes more negative.  This feature is illustrated by the nearly vertical dashed lines in Fig.~\ref{fig4} that connect $d_M$ values at constant $\bar \epsilon_{AB}$.  (Note that the correlation between the vertical shift in $d_M(S_{AB})$ and $\epsilon_\_$ breaks down for regions of the $\bar \epsilon_{AB}$-$\epsilon_\_$ plane with poor glass-forming ability for $\sigma_{BB}/\sigma_{AA}=1.05$.)

We have shown that changes in $d_M$ (and hence changes in the glass-forming ability) caused by changes in $\bar \epsilon_{AB}$ at fixed $\epsilon_\_$ couple strongly to the local chemical order. What is the particle-scale origin of the variations in $d_M$ caused by changes in $\epsilon_\_$ (at fixed $\bar \epsilon_{AB}$)?  In Fig.~\ref{fig5}, we show that the relative Voronoi volume difference 
$\langle \Delta V/V_A \rangle = \langle (V_A-V_B)/V_A \rangle$ 
(for systems where the larger particles have larger cohesive energy) increases as the cohesive energy difference $\epsilon_\_$ becomes more negative.  Decreasing the cohesive energy between the smaller B particles in the Lennard-Jones potential allows the average spacing between B particles to decrease below $2^{1/6} \sigma_{BB}$. (See Fig.~\ref{fig0}.) Thus, the smaller B particles occupy even less volume (relative to the larger A particles) as the cohesive energy of the less-cohesive B particles becomes smaller. For systems where the larger particles have 
larger cohesive energy, $\langle \Delta V \rangle > 0$ for all values of $\epsilon_\_ <0$. The variation of $\langle \Delta V/V_A \rangle$ versus $\epsilon_\_$ is similar for LJ systems with different diameter ratios and $\sigma_{BB}/\sigma_{AA} <1$; they simply differ by the size of the vertical shift in $\langle \Delta V/V_A \rangle$. 

For the binary Lennard-Jones system where the larger B particles have smaller cohesive energy, $\sigma_{BB}/\sigma_{AA}=1.05$, we use the same definition for the relative Voronoi volume difference, $\langle \Delta V/V_A \rangle$. In Fig.~\ref{fig5}, we show that for this system, $\langle \Delta V/V_A \rangle <0$ since the B particles are larger than the A particles. As $\epsilon_\_$ becomes more negative, $\langle \Delta V/V_A \rangle$ increases, approaching zero, since the larger B particles decrease their spacing due their decreased cohesive energy. (See Fig.~\ref{fig0}.) Thus, in all cases, decreases in $\epsilon_\_$ cause increases in $\langle \Delta V/V_A \rangle$ and $d_M$, which increases the glass-forming ability. 


As shown in Fig.~\ref{fig4}, for all diameter ratios and values of $\epsilon_\_$, the $d_M$ versus $S_{AB}$ curves possess hyperbolic shapes. We find that by shifting the curves in $S_{AB}$ and $d_M$ by $S_{AB}^0$ and $d_M^0$, respectively, (see Table~\ref{tab:table2}), they can be collapsed onto a single master curve.  In general, the shift in $S_{AB}$ is small, $|S_{AB}^0| \sim 0$, and $S_{AB}^0 < 0$. In contrast, $d_M^0 >0$ and it increases strongly with decreasing $\epsilon_\_$. 

\begin{figure}[t!]
\centering
 \includegraphics[width = \linewidth]{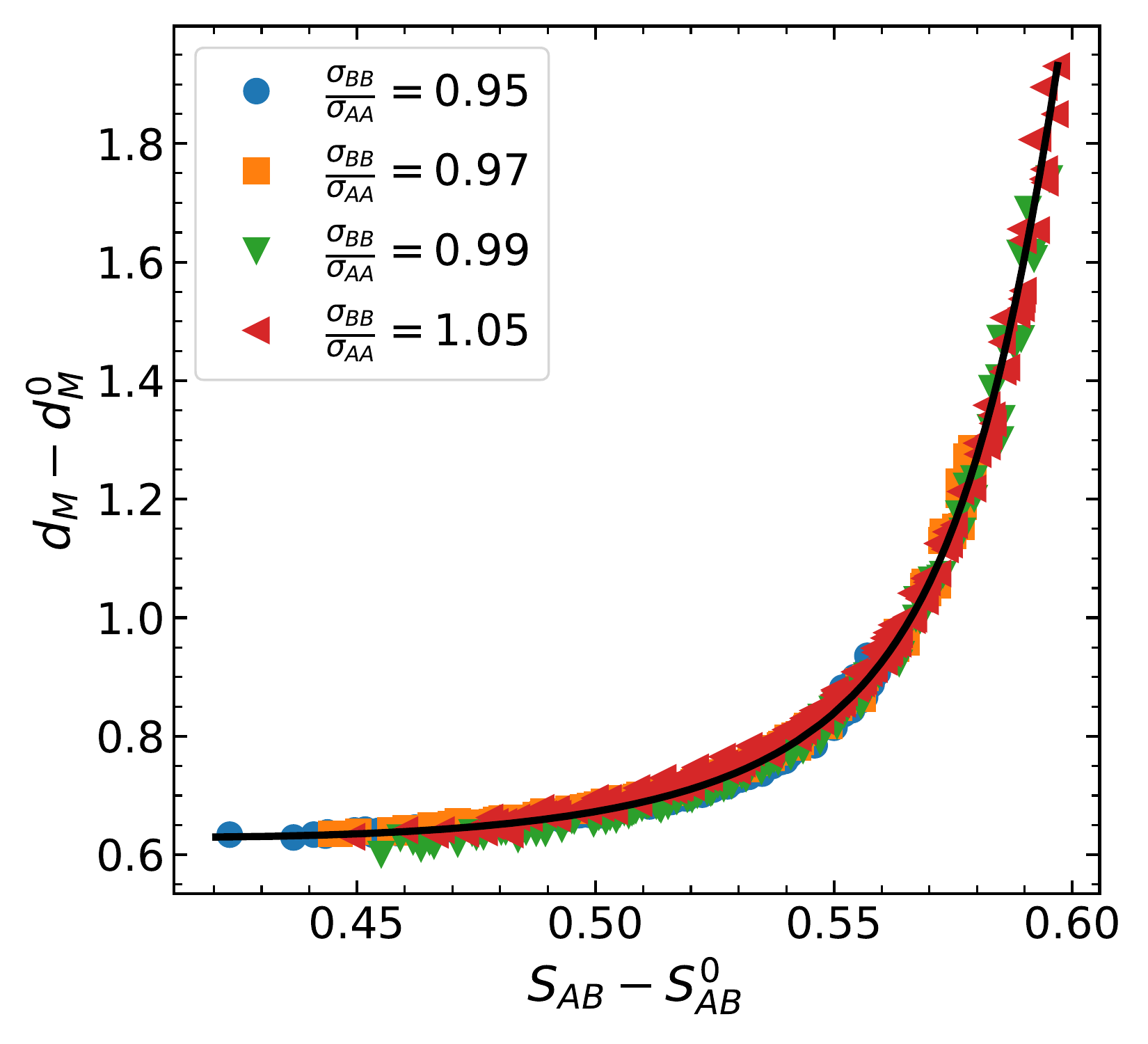}
 \caption{
Collapse of the data in Fig.~\ref{fig4} obtained by plotting $d_M-d_M^0$ versus $S_{AB}-S_{AB}^0$ for all four diameter ratios. The black solid line is the best fit of the data to the generalized hyperbolic form in Eq.~\ref{hyperbola} with R-squared $\approx 0.97$.
The parameters of the best-fit generalized hyperbola are $\theta =-130.6^\circ$, $A=0.382$, $B=0.390$, $S^c_{AB} - S^0_{AB} =0.592$, and $d^c_M-d^0_M = 0.442$.}
\label{fig6}
\end{figure}

In Fig.~\ref{fig6}, we show the collapse of all the data in Fig.~\ref{fig4} by plotting $d_M-d_M^0$ versus $S_{AB}-S_{AB}^0$. 
The master curve has the following general hyperbolic form: 
\begin{equation}
\label{hyperbola}
    (\bf \Delta-\Delta^c)^{\rm T}Q^{\rm T}\Lambda Q(\bf \Delta-\Delta^c)=1,
\end{equation}
where ${\bf \Delta} = (S_{AB}-S_{AB}^0,d_M - d_M^0)^{\rm T}$, 
${\bf \Delta^c}$ gives the rotation center, $\bf Q$ is the rotation matrix,
\begin{equation}
    Q=
\begin{bmatrix}
\cos\theta & -\sin\theta \\
\sin\theta & \cos\theta
\end{bmatrix},
\end{equation}
in the $(S_{AB}-S_{AB}^0)$-$(d_M-d_M^0)$ plane, and $\Lambda$ is the diagonal matrix with semi-axis lengths $A$ and $B$,
\begin{equation}
    \Lambda=
\begin{bmatrix}
\frac{1}{A^2} & 0 \\
0 & -\frac{1}{B^2}
\end{bmatrix}.
\end{equation}

The generalized hyperbolic form has five shape parameters, which take on the values $S_{ab}^c -S_{AB}^0 \approx 0.592$, $d_M^c-d_M^0 \approx 0.442$, $\theta \approx -130.6^{\circ}$, $A\approx 0.382$, and $B \approx 0.390$ for the master curve in Fig.~\ref{fig6}.  The R-squared value of the fit of all data in Fig.~\ref{fig6} to Eq.~\ref{hyperbola} is $\approx 0.97$. These results emphasize that the two dominant contributions to the glass-forming ability of binary Lennard-Jones systems are the local chemical order (where increases in $\bar \epsilon_{AB}$ cause increases in $S_{AB}$ and $d_M$) and the relative difference in the Voronoi volumes of the particles (where decreases in $\epsilon_\_$ cause increases in $d_M^0$). 

\begin{table*}[t!]
\caption{\label{tab:table2}%
The values of the shift parameters $S_{AB}^0$ and $d_M^0$ as a function of $\epsilon_\_$ that yield collapse of the data in Fig.~\ref{fig5} for all diameter ratios onto a master curve in the shape of a generalized hyperbola.}
\begin{ruledtabular}
\begin{tabular}{cccccccccccccc}
 &\multicolumn{2}{c}{$\sigma_{BB}/\sigma_{AA}=0.95$}
 &\multicolumn{2}{c}{$\sigma_{BB}/\sigma_{AA}=0.97$}
 &\multicolumn{2}{c}{$\sigma_{BB}/\sigma_{AA}=0.99$}
 &\multicolumn{2}{c}{$\sigma_{BB}/\sigma_{AA}=1.05$}\\
 $\epsilon_\_$
 &$S_{AB}^0$&$d_M^0$
 &$S_{AB}^0$&$d_M^0$
 &$S_{AB}^0$&$d_M^0$
 &$S_{AB}^0$&$d_M^0$\\ \hline
0      & 0.035 & 1.700 & 0.013 & 0.650 & -0.003 & -0.100 & 0      & 0      \\
-0.053 & 0.034 & 1.730 & 0.013 & 0.670 & -0.004 & -0.085 & -0.003 & -0.060 \\
-0.111 & 0.033 & 1.750 & 0.013 & 0.700 & 0      & -0.025 & -0.003 & -0.135 \\
-0.176 & 0.033 & 1.790 & 0.013 & 0.745 & 0.001  & 0.035  & -0.010 & -0.190 \\
-0.250 & 0.032 & 1.850 & 0.011 & 0.805 & -0.002 & 0.110  & -0.007 & -0.190 \\
-0.333 & 0.031 & 1.915 & 0.013 & 0.900 & 0.003  & 0.245  & -0.012 & -0.178 \\
-0.379 & 0.035 & 1.960 & 0.013 & 0.950 & 0.003  & 0.305  & -0.015 & -0.150 \\
-0.429 & 0.029 & 2.000 & 0.011 & 1.010 & 0.003  & 0.385  & -0.007 & -0.070 \\
-0.481 & 0.028 & 2.057 & 0.010 & 1.080 & 0.003  & 0.465  & -0.006 & -0.008 \\
-0.538 & 0.030 & 2.117 & 0.010 & 1.157 & 0.003  & 0.555  & -0.005 & 0.100  \\
-0.600 & 0.030 & 2.190 & 0.010 & 1.250 & 0.003  & 0.662  & -0.004 & 0.200  \\
-0.667 & -     & -     & -     & -     & 0.003  & 0.785  & -0.002 & 0.320 
\end{tabular}
\end{ruledtabular}
\end{table*}

\begin{figure*}[t!]
\centering
\includegraphics[width = 0.8\textwidth]{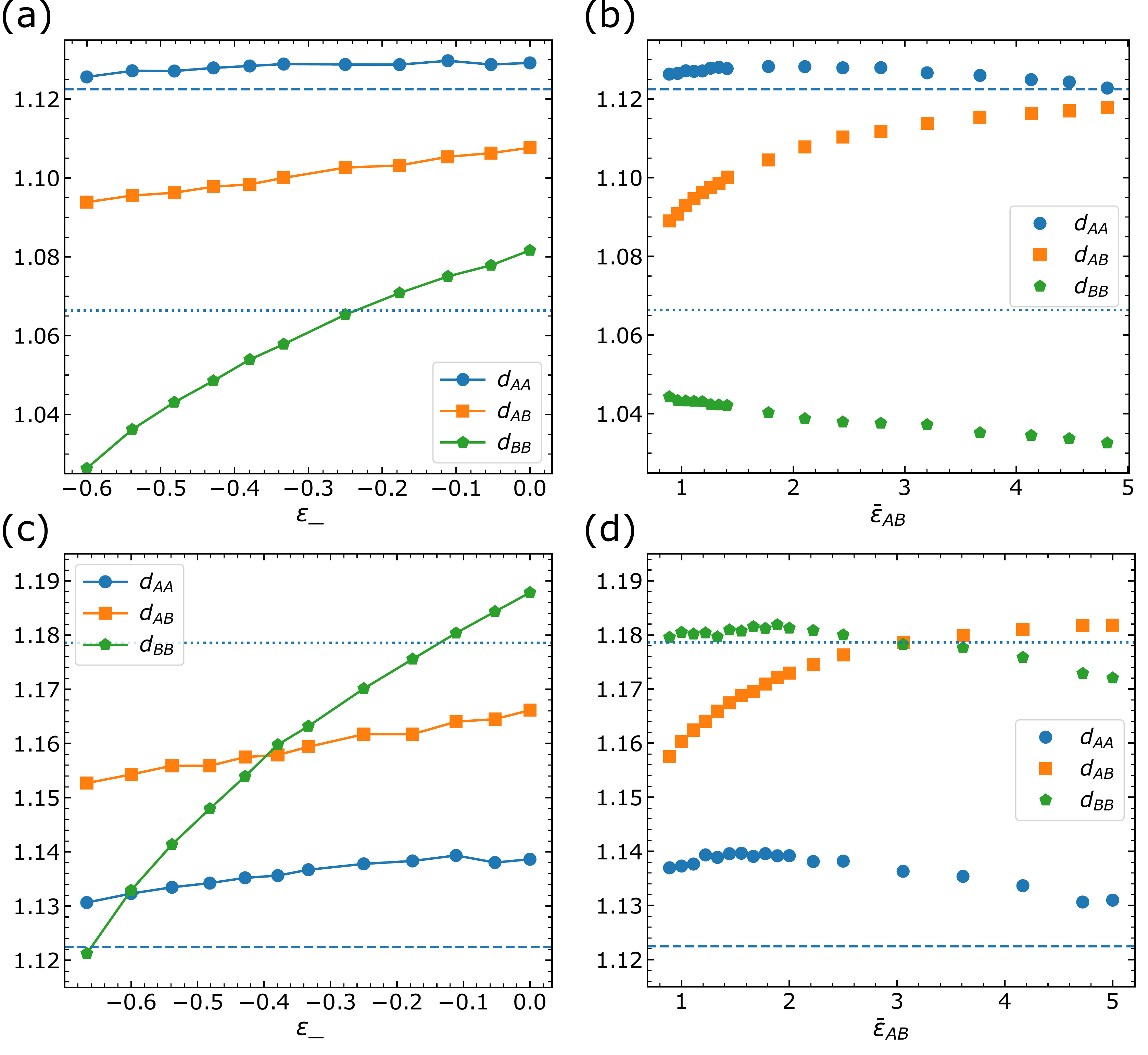}
\caption{Average pair separations, $d_{\alpha \beta}$, for samples generated at high cooling rates ($R =10^{-2}$) for diameter ratios
$\sigma_{BB}/\sigma_{AA}=0.95$  (panels (a) and (b)) and 
$\sigma_{BB}/\sigma_{AA}=1.05$  (panels (c) and (d)). 
In panels (a) and (c), we show $d_{\alpha \beta}$ as a function of $\epsilon_\_$ at $\bar \epsilon_{AB} \approx 1.20$. In 
panels (b) and (d), we show $d_{\alpha \beta}$ as a function of $\bar \epsilon_{AB}$ at $\epsilon_\_ = -0.111$. The dashed (dotted) lines give the positions of the minima in $V_{AA}$ ($V_{BB}$), i.e. $2^{1/6} \sigma_{AA}$ ($2^{1/6} \sigma_{BB}$).}
\label{fig7}
\end{figure*}

The asymmetry of the $\log_{10} R_c$ contours in the $\bar \epsilon_{AB}$-$\epsilon_\_$ plane between
systems for which the larger particle has larger cohesive energy and systems for which the larger particle has smaller cohesive energy can be further illustrated by comparing the average pair separations $d_{\alpha \beta}$ from the rapidly cooled samples. We show $d_{AA}$, $d_{AB}$, and $d_{BB}$ versus $\epsilon_\_$ in Fig.~\ref{fig7} (a) and versus $\bar \epsilon_{AB}$ in Fig.~\ref{fig7} (b) for $\sigma_{BB}/\sigma_{AA}=0.95$. We also show the corresponding data for $\sigma_{BB}/\sigma_{AA}=1.05$ in Fig.~\ref{fig7} (c) and (d).  

As expected, at fixed $\epsilon_\_$, $d_{AA}$ and $d_{BB}$ remain nearly constant as $\bar \epsilon_{AB}$ varies. In contrast, $d_{AB}$ increases by $\approx 0.03$ as $\bar \epsilon_{AB}$ increases by a factor of five, as shown in Fig.~\ref{fig7} (b) and (d). Thus, increases in the local chemical order and glass-forming ability caused by increases in $\bar \epsilon_{AB}$ are associated with increases in the interparticle spacing $d_{AB}$. In contrast, at fixed $\bar \epsilon_{AB}$,  $d_{BB}$ {\it decreases} strongly as $\epsilon_\_$ decreases from $0$ to $\approx -0.7$, whereas $d_{AA}$ and $d_{AB}$ remain nearly constant. (See Fig.~\ref{fig7} (a) and (c).)  For the system with diameter ratio $\sigma_{BB}/\sigma_{AA} =0.95$ (Fig.~\ref{fig7} (a)), the BB bonds shorten as a function of decreasing $\epsilon_\_$, while the AA bonds remain nearly constant with $d_{AA} > d_{BB}$. 
For the system with diameter ratio $\sigma_{BB}/\sigma_{AA} =1.05$, $d_{BB} > d_{AA}$ at $\epsilon_\_ \approx 0$. $d_{BB}$ decreases toward $d_{AA}$ with decreasing $\epsilon_\_$.  Only after $d_{BB} < d_{AA}$ upon further decreases in $\epsilon_\_$ does the glass-forming ability begin to increase. Thus, bond-shortening (where $d_{BB} < d_{AA}$ and $d_{BB} < 2^{1/6} \sigma_{BB}$) is a key factor that contributes to the glass-forming ability of binary Lennard-Jones systems. 

\section{IV. Conclusions and Future Directions}

In this work, we investigated the physical features that control the glass-forming ability of equimolar, binary mixtures.  We performed computational studies of binary Lennard-Jones systems, which enabled us to independently tune the particle sizes, cohesive and mixing energies and determine their effects on the glass-forming ability, as measured by the critical cooling rate $R_c$.  We showed four key results. First, the $\log_{10} R_c$ contours as a function of the normalized mixing energy $\bar \epsilon_{AB}$ and cohesive energy difference $\epsilon_\_$ are ellipsoidal in shape for all diameter ratios. 
The energetic parameters $\epsilon_\_$ and $\bar \epsilon_{AB}$ could be used to select element combinations using their  specific values to find best glass formers that can be realized from the periodic table.
Thus, the Mahalanobis distance $d_M$ from a given point in the $\bar \epsilon_{AB}$-$\epsilon_\_$ plane to the center of the ellipse determines the glass-forming ability of LJ systems.  In particular, we find that the data for all diameter ratios can be described by $\log_{10} R_c/R_0 = -d_M^2(\bar \epsilon_{AB},\epsilon_\_)$, where $R_0$ is a diameter ratio-dependent reference rate.  Second, by studying the structural properties of the low-temperature systems generated by rapid cooling, we showed that $d_M$ is controlled by the local chemical order $S_{AB}$ (i.e. deviations of the local composition from the globally averaged value) and relative Voronoi volume differences $\langle \Delta V/V_A\rangle$ between particles. Increases in $\bar \epsilon_{AB}$ cause increases in $S_{AB}$, $d_M$, and thus the glass-forming ability. Decreases in $\epsilon_\_$ (i.e. $\epsilon_\_$ becomes more negative) cause increases in $\langle \Delta V/V_A\rangle$, $d_M$, and the glass-forming ability. We find that by plotting $d_M - d_M^0$ versus $S_{AB} - S_{AB}^0$, where $d_M^0$ and $S_{AB}^0$ are shifts that depend on the diameter ratio and $\epsilon_\_$, we can collapse the data for all diameter ratios onto a generalized hyperbolic master curve. Third, we showed that LJ systems for which the larger particles have larger cohesive energy ($\sigma_{BB}/\sigma_{AA} < 1$ and $\epsilon_{BB}/\epsilon_{AA} < 1$) are better glass-formers than systems for which the larger particles have smaller cohesive energy ($\sigma_{BB}/\sigma_{AA} > 1$ and $\epsilon_{BB}/\epsilon_{AA} < 1)$. We illustrated this point by showing that the system with $\sigma_{BB}/\sigma_{AA} = 1.05$ possessed stronger bond shortening, where the typical interparticle separation $d_{BB}$ was much smaller than the location of the minimum in $V_{BB}$, than the system with $\sigma_{BB}/\sigma_{AA} = 0.95$. 

These results suggest several promising directions for future research aimed at understanding the GFA of particle mixtures. First, we focused here on equimolar, binary LJ systems. It will be important to understand whether our results hold over a wide range of compositions, $0 < f_B < 1$.  For example, is the result that LJ systems for which the larger particles have larger cohesive energy are the best glass formers also true in systems where the smaller particles are the majority particle species?  

As shown by recent combinatorial sputtering experiments, the GFA of metallic glasses depends on many physical features including the particle size differences, cohesive energy differences, mixing energy, as well as the elemental atomic symmetry (i.e. the crystalline symmetry, e.g. FCC, BCC, and HCP, that forms when the pure substance is crystallized)~\cite{ding_combinatorial_2014,li_how_2017,li_data-driven_2022}. In prior studies~\cite{gfa_2019,hu_glass_2020}, we investigated the effects of the cohesive energy differences, mixing energies, and atomic symmetry on the GFA of binary mixtures.  However, we did not include particle size differences in these prior studies. In future studies, we will carry out molecular dynamics simulations of the patchy particle model of binary mixtures with different atomic sizes to explore the coupling of atomic size differences to cohesive energy differences and differences in atomic symmetry in determining the GFA. 

Finally, our results highlight the importance of local chemical order and Voronoi volume differences in determining the GFA of binary LJ systems. In future studies, we will investigate the connection between the local chemical order, Voronoi volume differences, and GFA in ternary and quaternary LJ systems.  In addition, similar computational studies can be carried out using embedded atom method (EAM) potentials to determine whether the chemical order and Voronoi volume differences determine the glass-forming ability in more complex models of binary, ternary, and quaternary alloys. 

\section*{Acknowledgments}
The authors acknowledge support from NSF Grant Nos. DMR-1119826 (Y.-C.H.), CMMI-1901959 (C.O.), and CMMI-1463455 (M.S.). This work was supported by the High Performance Computing facilities operated by, and the staff of, the Yale Center for Research Computing.

\section*{Appendix}

In this Appendix, we investigate the local potential energy minima  for binary Lennard-Jones systems to better understand the GFA for different diameter ratios. In general, good glass-formers possess deeper local potential energy minima and higher barriers separating nearby minima. For these studies, we use the conjugate gradient energy minimization method to take of all of the low-temperature systems obtained by rapid cooling to the corresponding nearest potential energy minimum, or inherent structure~\cite{sastry_signatures_1998,debenedetti_supercooled_2001}. 
For each sample, we perform two sets of analyses. We first measure the total potential energy per particle of the inherent structures $E_{\rm IS}(\sigma_{BB}/\sigma_{AA})$ of the samples with the original values of $\epsilon_{AA}$, $\epsilon_{AB}$, $\epsilon_{BB}$, $\sigma_{AA}$, and $\sigma_{BB}$. We then change the sizes of the A and B particles within the original sample, keep the energetic parameters the same, and minimize the total potential energy to obtain $E_{\rm IS}^f$. 

\begin{figure*}[t!]
\centering
 \includegraphics[width = 0.9\textwidth]{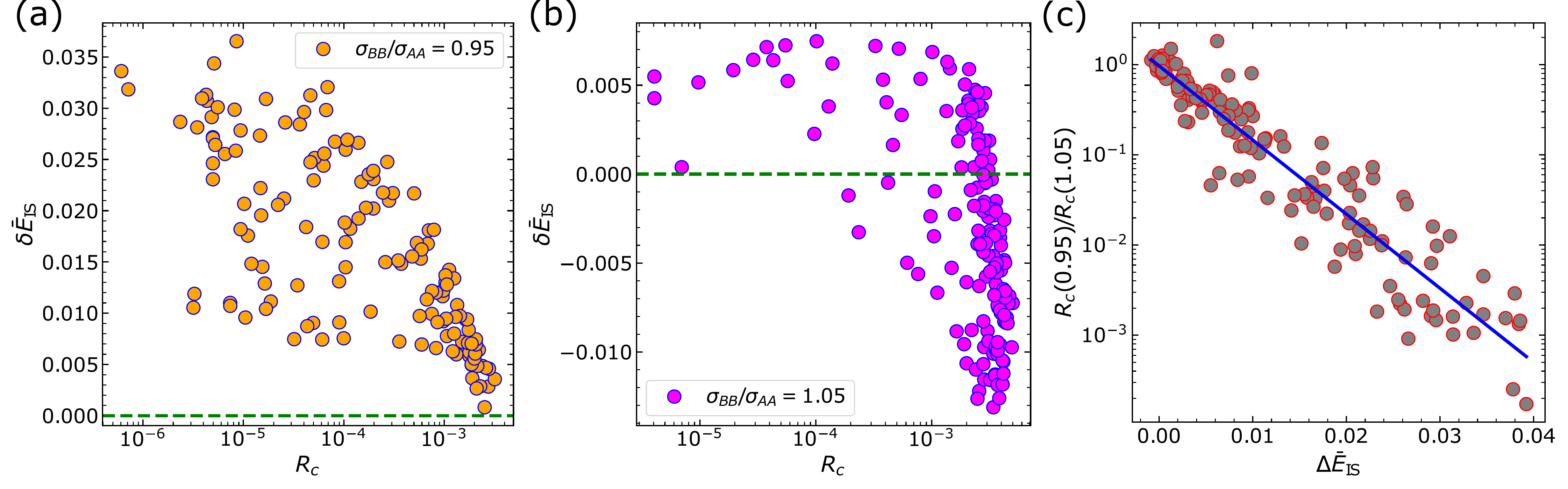}
 \caption{
(a) The relative difference in the average potential energy of the inherent structures, $\delta \bar E_{\rm IS} = (E_{\rm IS}(0.95) - E_{\rm IS}^f)/E_{\rm IS}(0.95)$, from systems at a given point in the $\bar \epsilon_{AB}$-$\epsilon_\_$ plane and $\sigma_{BB}/\sigma_{AA}=0.95$ and the same configurations with the particle size switched to $\sigma_{BB}/\sigma_{AA}=1.05$. (b) Similar plot of $\delta \bar E_{\rm IS}$ versus $R_c$ except the original samples have $\sigma_{BB}/\sigma_{AA}=1.05$ and the swapped samples have $\sigma_{BB}/\sigma_{AA}=0.95$. The horizontal dashed line indicates $\delta \bar E_{\rm IS}=0$. (c) $R_c(0.95)/R_c(1.05)$ (on a logarithmic scale) plotted versus the relative difference in the average potential energy of the inherent structures for systems with $\sigma_{BB}/\sigma_{AA}=0.95$ and $1.05$, $\Delta \bar E_{\rm IS} = (E_{\rm IS}(0.95) - E_{\rm IS}(1.05))/E_{\rm IS}(0.95)$. The solid line indicates $\log_{10} [R_c(0.95)/R_c(1.05)] = A \Delta \bar E_{\rm IS} - B$, where $A$ and $B$ are constants. }
\label{figA1}
\end{figure*}

In Fig.~\ref{figA1} (a) and (b),  we show the relative change in the total potential energy of the inherent structures,
\begin{equation}
\delta \bar E_{\rm IS} = \frac{E_{\rm IS}(\sigma_{BB}/\sigma_{AA}) - E_{\rm IS}^f}{E_{\rm IS}(\sigma_{BB}/\sigma_{AA})},
\end{equation}
as a function of $R_c$, where $E_{\rm IS}(\sigma_{BB}/\sigma_{AA}) < 0$.  $\delta \bar E_{\rm IS} >0$ indicates that the inherent structures for systems with a given diameter ratio of the smaller to the larger particles, $\sigma_{BB}/\sigma_{AA} <1$ with $\epsilon_{BB}/\epsilon_{AA} <1$, is more stable than those configurations with the sizes of the particles switched. 
As shown in Fig.~\ref{figA1} (a), for LJ systems with $\sigma_{BB}/\sigma_{AA}=0.95$, $\delta \bar E_{\rm IS} >0$, which indicates that LJ systems with $\sigma_{BB}/\sigma_{AA}=0.95$ are more stable than similar configurations with $\sigma_{BB}/\sigma_{AA}=1.05$ and the same energetic parameters. 

In contrast, in Fig.~\ref{figA1} (b), we show that for LJ systems with $\sigma_{BB}/\sigma_{AA}=1.05$, most of the data satisfies $\delta \bar E_{\rm IS} <0$ or $\delta \bar E_{\rm IS} \sim 0$ when $\delta \bar E_{\rm IS} >0$.  Thus, LJ systems for which the smaller particles have larger cohesive energy possess inherent structures that are typically less stable than those for the opposite case (where the larger particles have larger cohesive energy).

We also compare the logarithmic differences in the critical cooling rates, $\log_{10} R_c(0.95)/R_c(1.05)$, to the relative difference in the inherent structure energy,
\begin{equation}
\Delta \bar E_{\rm IS} = \frac{E_{\rm IS}(0.95) - E_{\rm IS} (1.05)}{E_{\rm IS}(0.95)},
\end{equation}
for LJ systems with $\sigma_{BB}/\sigma_{AA} =0.95$ and $1.05$ in Fig.~\ref{figA1} (c).  We find an approximate linear correlation between $\log_{10} [R_c(0.95)/R_c(1.05)]$ and $\Delta \bar E_{\rm IS}$, 
which indicates Arrhenius dependence of the critical cooling rate on the inherent structure energy. Thus, LJ systems with lower inherent structure energy possess better glass-forming ability. 


\begin{thebibliography}{32}%
\makeatletter
\providecommand \@ifxundefined [1]{%
 \@ifx{#1\undefined}
}%
\providecommand \@ifnum [1]{%
 \ifnum #1\expandafter \@firstoftwo
 \else \expandafter \@secondoftwo
 \fi
}%
\providecommand \@ifx [1]{%
 \ifx #1\expandafter \@firstoftwo
 \else \expandafter \@secondoftwo
 \fi
}%
\providecommand \natexlab [1]{#1}%
\providecommand \enquote  [1]{``#1''}%
\providecommand \bibnamefont  [1]{#1}%
\providecommand \bibfnamefont [1]{#1}%
\providecommand \citenamefont [1]{#1}%
\providecommand \href@noop [0]{\@secondoftwo}%
\providecommand \href [0]{\begingroup \@sanitize@url \@href}%
\providecommand \@href[1]{\@@startlink{#1}\@@href}%
\providecommand \@@href[1]{\endgroup#1\@@endlink}%
\providecommand \@sanitize@url [0]{\catcode `\\12\catcode `\$12\catcode
  `\&12\catcode `\#12\catcode `\^12\catcode `\_12\catcode `\%12\relax}%
\providecommand \@@startlink[1]{}%
\providecommand \@@endlink[0]{}%
\providecommand \url  [0]{\begingroup\@sanitize@url \@url }%
\providecommand \@url [1]{\endgroup\@href {#1}{\urlprefix }}%
\providecommand \urlprefix  [0]{URL }%
\providecommand \Eprint [0]{\href }%
\providecommand \doibase [0]{https://doi.org/}%
\providecommand \selectlanguage [0]{\@gobble}%
\providecommand \bibinfo  [0]{\@secondoftwo}%
\providecommand \bibfield  [0]{\@secondoftwo}%
\providecommand \translation [1]{[#1]}%
\providecommand \BibitemOpen [0]{}%
\providecommand \bibitemStop [0]{}%
\providecommand \bibitemNoStop [0]{.\EOS\space}%
\providecommand \EOS [0]{\spacefactor3000\relax}%
\providecommand \BibitemShut  [1]{\csname bibitem#1\endcsname}%
\let\auto@bib@innerbib\@empty
\bibitem [{\citenamefont {Zhong}\ \emph {et~al.}(2014)\citenamefont {Zhong},
  \citenamefont {Wang}, \citenamefont {Sheng}, \citenamefont {Zhang},\ and\
  \citenamefont {Mao}}]{zhong_formation_2014}%
  \BibitemOpen
  \bibfield  {author} {\bibinfo {author} {\bibfnamefont {L.}~\bibnamefont
  {Zhong}}, \bibinfo {author} {\bibfnamefont {J.}~\bibnamefont {Wang}},
  \bibinfo {author} {\bibfnamefont {H.}~\bibnamefont {Sheng}}, \bibinfo
  {author} {\bibfnamefont {Z.}~\bibnamefont {Zhang}},\ and\ \bibinfo {author}
  {\bibfnamefont {S.~X.}\ \bibnamefont {Mao}},\ }\bibfield  {title} {\bibinfo
  {title} {Formation of monatomic metallic glasses through ultrafast liquid
  quenching},\ }\href@noop {} {\bibfield  {journal} {\bibinfo  {journal}
  {Nature}\ }\textbf {\bibinfo {volume} {512}},\ \bibinfo {pages} {177}
  (\bibinfo {year} {2014})}\BibitemShut {NoStop}%
\bibitem [{\citenamefont {Wang}\ \emph {et~al.}(2004)\citenamefont {Wang},
  \citenamefont {Dong},\ and\ \citenamefont {Shek}}]{wang_bulk_2004}%
  \BibitemOpen
  \bibfield  {author} {\bibinfo {author} {\bibfnamefont {W.~H.}\ \bibnamefont
  {Wang}}, \bibinfo {author} {\bibfnamefont {C.}~\bibnamefont {Dong}},\ and\
  \bibinfo {author} {\bibfnamefont {C.~H.}\ \bibnamefont {Shek}},\ }\bibfield
  {title} {\bibinfo {title} {Bulk metallic glasses},\ }\href@noop {} {\bibfield
   {journal} {\bibinfo  {journal} {Mater. Sci. Eng.: R Rep.}\ }\textbf
  {\bibinfo {volume} {44}},\ \bibinfo {pages} {45} (\bibinfo {year}
  {2004})}\BibitemShut {NoStop}%
\bibitem [{\citenamefont {Takeuchi}\ and\ \citenamefont
  {Inoue}(2005)}]{takeuchi_classification_2005}%
  \BibitemOpen
  \bibfield  {author} {\bibinfo {author} {\bibfnamefont {A.}~\bibnamefont
  {Takeuchi}}\ and\ \bibinfo {author} {\bibfnamefont {A.}~\bibnamefont
  {Inoue}},\ }\bibfield  {title} {\bibinfo {title} {Classification of bulk
  metallic glasses by atomic size difference, heat of mixing and period of
  constituent elements and its application to characterization of the main
  alloying element},\ }\href@noop {} {\bibfield  {journal} {\bibinfo  {journal}
  {Mater. Trans.}\ }\textbf {\bibinfo {volume} {46}},\ \bibinfo {pages} {2817}
  (\bibinfo {year} {2005})}\BibitemShut {NoStop}%
\bibitem [{\citenamefont {Lu}\ and\ \citenamefont {Liu}(2002)}]{lu_new_2002}%
  \BibitemOpen
  \bibfield  {author} {\bibinfo {author} {\bibfnamefont {Z.~P.}\ \bibnamefont
  {Lu}}\ and\ \bibinfo {author} {\bibfnamefont {C.~T.}\ \bibnamefont {Liu}},\
  }\bibfield  {title} {\bibinfo {title} {A new glass-forming ability criterion
  for bulk metallic glasses},\ }\href@noop {} {\bibfield  {journal} {\bibinfo
  {journal} {Acta Mater.}\ }\textbf {\bibinfo {volume} {50}},\ \bibinfo {pages}
  {3501} (\bibinfo {year} {2002})}\BibitemShut {NoStop}%
\bibitem [{\citenamefont {Johnson}\ \emph {et~al.}(2016)\citenamefont
  {Johnson}, \citenamefont {Na},\ and\ \citenamefont
  {Demetriou}}]{johnson_quantifying_2016}%
  \BibitemOpen
  \bibfield  {author} {\bibinfo {author} {\bibfnamefont {W.~L.}\ \bibnamefont
  {Johnson}}, \bibinfo {author} {\bibfnamefont {J.~H.}\ \bibnamefont {Na}},\
  and\ \bibinfo {author} {\bibfnamefont {M.~D.}\ \bibnamefont {Demetriou}},\
  }\bibfield  {title} {\bibinfo {title} {Quantifying the origin of metallic
  glass formation},\ }\href@noop {} {\bibfield  {journal} {\bibinfo  {journal}
  {Nat. Commun.}\ }\textbf {\bibinfo {volume} {7}},\ \bibinfo {pages} {10313}
  (\bibinfo {year} {2016})}\BibitemShut {NoStop}%
\bibitem [{\citenamefont {Auer}\ and\ \citenamefont
  {Frenkel}(2001)}]{auer_suppression_2001}%
  \BibitemOpen
  \bibfield  {author} {\bibinfo {author} {\bibfnamefont {S.}~\bibnamefont
  {Auer}}\ and\ \bibinfo {author} {\bibfnamefont {D.}~\bibnamefont {Frenkel}},\
  }\bibfield  {title} {\bibinfo {title} {Suppression of crystal nucleation in
  polydisperse colloids due to increase of the surface free energy},\
  }\href@noop {} {\bibfield  {journal} {\bibinfo  {journal} {Nature}\ }\textbf
  {\bibinfo {volume} {413}},\ \bibinfo {pages} {711} (\bibinfo {year}
  {2001})}\BibitemShut {NoStop}%
\bibitem [{\citenamefont {Tanaka}(2012)}]{tanaka_bond_2012}%
  \BibitemOpen
  \bibfield  {author} {\bibinfo {author} {\bibfnamefont {H.}~\bibnamefont
  {Tanaka}},\ }\bibfield  {title} {\bibinfo {title} {Bond orientational order
  in liquids: {Towards} a unified description of water-like anomalies,
  liquid-liquid transition, glass transition, and crystallization},\
  }\href@noop {} {\bibfield  {journal} {\bibinfo  {journal} {Eur. Phys. J. E}\
  }\textbf {\bibinfo {volume} {35}},\ \bibinfo {pages} {1} (\bibinfo {year}
  {2012})}\BibitemShut {NoStop}%
\bibitem [{\citenamefont {Demetriou}\ \emph {et~al.}(2011)\citenamefont
  {Demetriou}, \citenamefont {Launey}, \citenamefont {Garrett}, \citenamefont
  {Schramm}, \citenamefont {Hofmann}, \citenamefont {Johnson},\ and\
  \citenamefont {Ritchie}}]{demetriou_damage-tolerant_2011}%
  \BibitemOpen
  \bibfield  {author} {\bibinfo {author} {\bibfnamefont {M.~D.}\ \bibnamefont
  {Demetriou}}, \bibinfo {author} {\bibfnamefont {M.~E.}\ \bibnamefont
  {Launey}}, \bibinfo {author} {\bibfnamefont {G.}~\bibnamefont {Garrett}},
  \bibinfo {author} {\bibfnamefont {J.~P.}\ \bibnamefont {Schramm}}, \bibinfo
  {author} {\bibfnamefont {D.~C.}\ \bibnamefont {Hofmann}}, \bibinfo {author}
  {\bibfnamefont {W.~L.}\ \bibnamefont {Johnson}},\ and\ \bibinfo {author}
  {\bibfnamefont {R.~O.}\ \bibnamefont {Ritchie}},\ }\bibfield  {title}
  {\bibinfo {title} {A damage-tolerant glass},\ }\href@noop {} {\bibfield
  {journal} {\bibinfo  {journal} {Nat. Mater.}\ }\textbf {\bibinfo {volume}
  {10}},\ \bibinfo {pages} {123} (\bibinfo {year} {2011})}\BibitemShut
  {NoStop}%
\bibitem [{\citenamefont {Li}\ \emph {et~al.}(2019)\citenamefont {Li},
  \citenamefont {Zhao}, \citenamefont {Lu}, \citenamefont {Hirata},
  \citenamefont {Wen}, \citenamefont {Bai}, \citenamefont {Chen}, \citenamefont
  {Schroers}, \citenamefont {Liu},\ and\ \citenamefont
  {Wang}}]{li_high-temperature_2019}%
  \BibitemOpen
  \bibfield  {author} {\bibinfo {author} {\bibfnamefont {M.-X.}\ \bibnamefont
  {Li}}, \bibinfo {author} {\bibfnamefont {S.-F.}\ \bibnamefont {Zhao}},
  \bibinfo {author} {\bibfnamefont {Z.}~\bibnamefont {Lu}}, \bibinfo {author}
  {\bibfnamefont {A.}~\bibnamefont {Hirata}}, \bibinfo {author} {\bibfnamefont
  {P.}~\bibnamefont {Wen}}, \bibinfo {author} {\bibfnamefont {H.-Y.}\
  \bibnamefont {Bai}}, \bibinfo {author} {\bibfnamefont {M.}~\bibnamefont
  {Chen}}, \bibinfo {author} {\bibfnamefont {J.}~\bibnamefont {Schroers}},
  \bibinfo {author} {\bibfnamefont {Y.}~\bibnamefont {Liu}},\ and\ \bibinfo
  {author} {\bibfnamefont {W.-H.}\ \bibnamefont {Wang}},\ }\bibfield  {title}
  {\bibinfo {title} {High-temperature bulk metallic glasses developed by
  combinatorial methods},\ }\href@noop {} {\bibfield  {journal} {\bibinfo
  {journal} {Nature}\ }\textbf {\bibinfo {volume} {569}},\ \bibinfo {pages}
  {99} (\bibinfo {year} {2019})}\BibitemShut {NoStop}%
\bibitem [{\citenamefont {Schroers}(2010)}]{schroers_processing_2010}%
  \BibitemOpen
  \bibfield  {author} {\bibinfo {author} {\bibfnamefont {J.}~\bibnamefont
  {Schroers}},\ }\bibfield  {title} {\bibinfo {title} {Processing of bulk
  metallic glass},\ }\href@noop {} {\bibfield  {journal} {\bibinfo  {journal}
  {Adv. Mater.}\ }\textbf {\bibinfo {volume} {22}},\ \bibinfo {pages} {1566}
  (\bibinfo {year} {2010})}\BibitemShut {NoStop}%
\bibitem [{\citenamefont {Ashby}\ and\ \citenamefont
  {Greer}(2006)}]{ashby_metallic_2006}%
  \BibitemOpen
  \bibfield  {author} {\bibinfo {author} {\bibfnamefont {M.~F.}\ \bibnamefont
  {Ashby}}\ and\ \bibinfo {author} {\bibfnamefont {A.~L.}\ \bibnamefont
  {Greer}},\ }\bibfield  {title} {\bibinfo {title} {Metallic glasses as
  structural materials},\ }\href@noop {} {\bibfield  {journal} {\bibinfo
  {journal} {Scr. Mater.}\ }\textbf {\bibinfo {volume} {54}},\ \bibinfo {pages}
  {321} (\bibinfo {year} {2006})}\BibitemShut {NoStop}%
\bibitem [{\citenamefont {Johnson}(2015)}]{johnson_is_2015}%
  \BibitemOpen
  \bibfield  {author} {\bibinfo {author} {\bibfnamefont {W.}~\bibnamefont
  {Johnson}},\ }\bibfield  {title} {\bibinfo {title} {Is metallic glass poised
  to come of age?},\ }\href@noop {} {\bibfield  {journal} {\bibinfo  {journal}
  {Nat. Mater.}\ }\textbf {\bibinfo {volume} {14}},\ \bibinfo {pages} {553}
  (\bibinfo {year} {2015})}\BibitemShut {NoStop}%
\bibitem [{\citenamefont {Ding}\ \emph {et~al.}(2014)\citenamefont {Ding},
  \citenamefont {Liu}, \citenamefont {Li}, \citenamefont {Liu}, \citenamefont
  {Sohn}, \citenamefont {Walker},\ and\ \citenamefont
  {Schroers}}]{ding_combinatorial_2014}%
  \BibitemOpen
  \bibfield  {author} {\bibinfo {author} {\bibfnamefont {S.}~\bibnamefont
  {Ding}}, \bibinfo {author} {\bibfnamefont {Y.}~\bibnamefont {Liu}}, \bibinfo
  {author} {\bibfnamefont {Y.}~\bibnamefont {Li}}, \bibinfo {author}
  {\bibfnamefont {Z.}~\bibnamefont {Liu}}, \bibinfo {author} {\bibfnamefont
  {S.}~\bibnamefont {Sohn}}, \bibinfo {author} {\bibfnamefont {F.~J.}\
  \bibnamefont {Walker}},\ and\ \bibinfo {author} {\bibfnamefont
  {J.}~\bibnamefont {Schroers}},\ }\bibfield  {title} {\bibinfo {title}
  {Combinatorial development of bulk metallic glasses},\ }\href@noop {}
  {\bibfield  {journal} {\bibinfo  {journal} {Nat. Mater.}\ }\textbf {\bibinfo
  {volume} {13}},\ \bibinfo {pages} {494} (\bibinfo {year} {2014})}\BibitemShut
  {NoStop}%
\bibitem [{\citenamefont {Li}\ \emph {et~al.}(2017)\citenamefont {Li},
  \citenamefont {Zhao}, \citenamefont {Liu}, \citenamefont {Gong},\ and\
  \citenamefont {Schroers}}]{li_how_2017}%
  \BibitemOpen
  \bibfield  {author} {\bibinfo {author} {\bibfnamefont {Y.}~\bibnamefont
  {Li}}, \bibinfo {author} {\bibfnamefont {S.}~\bibnamefont {Zhao}}, \bibinfo
  {author} {\bibfnamefont {Y.}~\bibnamefont {Liu}}, \bibinfo {author}
  {\bibfnamefont {P.}~\bibnamefont {Gong}},\ and\ \bibinfo {author}
  {\bibfnamefont {J.}~\bibnamefont {Schroers}},\ }\bibfield  {title} {\bibinfo
  {title} {How many bulk metallic glasses are there?},\ }\href@noop {}
  {\bibfield  {journal} {\bibinfo  {journal} {ACS Comb. Sci.}\ }\textbf
  {\bibinfo {volume} {19}},\ \bibinfo {pages} {687} (\bibinfo {year}
  {2017})}\BibitemShut {NoStop}%
\bibitem [{\citenamefont {Li}\ \emph {et~al.}(2022)\citenamefont {Li},
  \citenamefont {Sun}, \citenamefont {Wang}, \citenamefont {Hu}, \citenamefont
  {Sohn}, \citenamefont {Schroers}, \citenamefont {Wang},\ and\ \citenamefont
  {Liu}}]{li_data-driven_2022}%
  \BibitemOpen
  \bibfield  {author} {\bibinfo {author} {\bibfnamefont {M.-X.}\ \bibnamefont
  {Li}}, \bibinfo {author} {\bibfnamefont {Y.-T.}\ \bibnamefont {Sun}},
  \bibinfo {author} {\bibfnamefont {C.}~\bibnamefont {Wang}}, \bibinfo {author}
  {\bibfnamefont {L.-W.}\ \bibnamefont {Hu}}, \bibinfo {author} {\bibfnamefont
  {S.}~\bibnamefont {Sohn}}, \bibinfo {author} {\bibfnamefont {J.}~\bibnamefont
  {Schroers}}, \bibinfo {author} {\bibfnamefont {W.-H.}\ \bibnamefont {Wang}},\
  and\ \bibinfo {author} {\bibfnamefont {Y.-H.}\ \bibnamefont {Liu}},\
  }\bibfield  {title} {\bibinfo {title} {Data-driven discovery of a universal
  indicator for metallic glass forming ability},\ }\href@noop {} {\bibfield
  {journal} {\bibinfo  {journal} {Nat. Mater.}\ }\textbf {\bibinfo {volume}
  {21}},\ \bibinfo {pages} {165} (\bibinfo {year} {2022})}\BibitemShut
  {NoStop}%
\bibitem [{\citenamefont {Nishiyama}\ and\ \citenamefont
  {Inoue}(2002)}]{nishiyama_glass-forming_2002}%
  \BibitemOpen
  \bibfield  {author} {\bibinfo {author} {\bibfnamefont {N.}~\bibnamefont
  {Nishiyama}}\ and\ \bibinfo {author} {\bibfnamefont {A.}~\bibnamefont
  {Inoue}},\ }\bibfield  {title} {\bibinfo {title} {Glass-forming ability of
  {Pd42}.{5Cu30Ni7}.{5P20} alloy with a low critical cooling rate of 0.067
  {K}/s},\ }\href@noop {} {\bibfield  {journal} {\bibinfo  {journal} {Appl.
  Phys. Lett.}\ }\textbf {\bibinfo {volume} {80}},\ \bibinfo {pages} {568}
  (\bibinfo {year} {2002})}\BibitemShut {NoStop}%
\bibitem [{\citenamefont {Colvin}(2001)}]{colvin_opals_2001}%
  \BibitemOpen
  \bibfield  {author} {\bibinfo {author} {\bibfnamefont {V.~L.}\ \bibnamefont
  {Colvin}},\ }\bibfield  {title} {\bibinfo {title} {From opals to optics:
  colloidal photonic crystals},\ }\href@noop {} {\bibfield  {journal} {\bibinfo
   {journal} {MRS Bull.}\ }\textbf {\bibinfo {volume} {26}},\ \bibinfo {pages}
  {637} (\bibinfo {year} {2001})}\BibitemShut {NoStop}%
\bibitem [{\citenamefont {Fudouzi}\ and\ \citenamefont
  {Xia}(2003)}]{fudouzi_colloidal_2003}%
  \BibitemOpen
  \bibfield  {author} {\bibinfo {author} {\bibfnamefont {H.}~\bibnamefont
  {Fudouzi}}\ and\ \bibinfo {author} {\bibfnamefont {Y.}~\bibnamefont {Xia}},\
  }\bibfield  {title} {\bibinfo {title} {Colloidal crystals with tunable colors
  and their use as photonic papers},\ }\href@noop {} {\bibfield  {journal}
  {\bibinfo  {journal} {Langmuir}\ }\textbf {\bibinfo {volume} {19}},\ \bibinfo
  {pages} {9653} (\bibinfo {year} {2003})}\BibitemShut {NoStop}%
\bibitem [{\citenamefont {Goerlitzer}\ \emph {et~al.}(2018)\citenamefont
  {Goerlitzer}, \citenamefont {Klupp~Taylor},\ and\ \citenamefont
  {Vogel}}]{goerlitzer_bioinspired_2018}%
  \BibitemOpen
  \bibfield  {author} {\bibinfo {author} {\bibfnamefont {E.~S.~A.}\
  \bibnamefont {Goerlitzer}}, \bibinfo {author} {\bibfnamefont {R.~N.}\
  \bibnamefont {Klupp~Taylor}},\ and\ \bibinfo {author} {\bibfnamefont
  {N.}~\bibnamefont {Vogel}},\ }\bibfield  {title} {\bibinfo {title}
  {Bioinspired photonic pigments from colloidal self-assembly},\ }\href@noop {}
  {\bibfield  {journal} {\bibinfo  {journal} {Adv. Mater.}\ }\textbf {\bibinfo
  {volume} {30}},\ \bibinfo {pages} {1706654} (\bibinfo {year}
  {2018})}\BibitemShut {NoStop}%
\bibitem [{\citenamefont {Zhang}\ \emph {et~al.}(2013)\citenamefont {Zhang},
  \citenamefont {Wang}, \citenamefont {Papanikolaou}, \citenamefont {Liu},
  \citenamefont {Schroers}, \citenamefont {Shattuck}, \citenamefont {apos},\
  and\ \citenamefont {Hern}}]{kai_zhang_computational_2013}%
  \BibitemOpen
  \bibfield  {author} {\bibinfo {author} {\bibfnamefont {K.}~\bibnamefont
  {Zhang}}, \bibinfo {author} {\bibfnamefont {M.}~\bibnamefont {Wang}},
  \bibinfo {author} {\bibfnamefont {S.}~\bibnamefont {Papanikolaou}}, \bibinfo
  {author} {\bibfnamefont {Y.}~\bibnamefont {Liu}}, \bibinfo {author}
  {\bibfnamefont {J.}~\bibnamefont {Schroers}}, \bibinfo {author}
  {\bibfnamefont {M.~D.}\ \bibnamefont {Shattuck}}, \bibinfo {author}
  {\bibnamefont {apos}},\ and\ \bibinfo {author} {\bibfnamefont {C.~S.}\
  \bibnamefont {Hern}},\ }\bibfield  {title} {\bibinfo {title} {Computational
  studies of the glass-forming ability of model bulk metallic glasses},\
  }\href@noop {} {\bibfield  {journal} {\bibinfo  {journal} {J. Chem. Phys.}\
  }\textbf {\bibinfo {volume} {139}},\ \bibinfo {pages} {124503} (\bibinfo
  {year} {2013})}\BibitemShut {NoStop}%
\bibitem [{\citenamefont {Hu}\ \emph {et~al.}(2020)\citenamefont {Hu},
  \citenamefont {Zhang}, \citenamefont {Kube}, \citenamefont {Schroers},
  \citenamefont {Shattuck},\ and\ \citenamefont {O'Hern}}]{hu_glass_2020}%
  \BibitemOpen
  \bibfield  {author} {\bibinfo {author} {\bibfnamefont {Y.-C.}\ \bibnamefont
  {Hu}}, \bibinfo {author} {\bibfnamefont {K.}~\bibnamefont {Zhang}}, \bibinfo
  {author} {\bibfnamefont {S.~A.}\ \bibnamefont {Kube}}, \bibinfo {author}
  {\bibfnamefont {J.}~\bibnamefont {Schroers}}, \bibinfo {author}
  {\bibfnamefont {M.~D.}\ \bibnamefont {Shattuck}},\ and\ \bibinfo {author}
  {\bibfnamefont {C.~S.}\ \bibnamefont {O'Hern}},\ }\bibfield  {title}
  {\bibinfo {title} {Glass formation in binary alloys with different atomic
  symmetries},\ }\href@noop {} {\bibfield  {journal} {\bibinfo  {journal}
  {Phys. Rev. Mater.}\ }\textbf {\bibinfo {volume} {4}},\ \bibinfo {pages}
  {105602} (\bibinfo {year} {2020})}\BibitemShut {NoStop}%
\bibitem [{\citenamefont {Zhang}\ \emph {et~al.}(2014)\citenamefont {Zhang},
  \citenamefont {Smith}, \citenamefont {Wang}, \citenamefont {Liu},
  \citenamefont {Schroers}, \citenamefont {Shattuck},\ and\ \citenamefont
  {O'Hern}}]{zhang_connection_2014}%
  \BibitemOpen
  \bibfield  {author} {\bibinfo {author} {\bibfnamefont {K.}~\bibnamefont
  {Zhang}}, \bibinfo {author} {\bibfnamefont {W.~W.}\ \bibnamefont {Smith}},
  \bibinfo {author} {\bibfnamefont {M.}~\bibnamefont {Wang}}, \bibinfo {author}
  {\bibfnamefont {Y.}~\bibnamefont {Liu}}, \bibinfo {author} {\bibfnamefont
  {J.}~\bibnamefont {Schroers}}, \bibinfo {author} {\bibfnamefont {M.~D.}\
  \bibnamefont {Shattuck}},\ and\ \bibinfo {author} {\bibfnamefont {C.~S.}\
  \bibnamefont {O'Hern}},\ }\bibfield  {title} {\bibinfo {title} {Connection
  between the packing efficiency of binary hard spheres and the glass-forming
  ability of bulk metallic glasses},\ }\href@noop {} {\bibfield  {journal}
  {\bibinfo  {journal} {Phys. Rev. E}\ }\textbf {\bibinfo {volume} {90}},\
  \bibinfo {pages} {032311} (\bibinfo {year} {2014})}\BibitemShut {NoStop}%
\bibitem [{\citenamefont {Zhang}\ \emph {et~al.}(2015)\citenamefont {Zhang},
  \citenamefont {Fan}, \citenamefont {Liu}, \citenamefont {Schroers},
  \citenamefont {Shattuck},\ and\ \citenamefont {O'Hern}}]{zhang_beyond_2015}%
  \BibitemOpen
  \bibfield  {author} {\bibinfo {author} {\bibfnamefont {K.}~\bibnamefont
  {Zhang}}, \bibinfo {author} {\bibfnamefont {M.}~\bibnamefont {Fan}}, \bibinfo
  {author} {\bibfnamefont {Y.}~\bibnamefont {Liu}}, \bibinfo {author}
  {\bibfnamefont {J.}~\bibnamefont {Schroers}}, \bibinfo {author}
  {\bibfnamefont {M.~D.}\ \bibnamefont {Shattuck}},\ and\ \bibinfo {author}
  {\bibfnamefont {C.~S.}\ \bibnamefont {O'Hern}},\ }\bibfield  {title}
  {\bibinfo {title} {Beyond packing of hard spheres: {The} effects of core
  softness, non-additivity, intermediate-range repulsion, and many-body
  interactions on the glass-forming ability of bulk metallic glasses},\
  }\href@noop {} {\bibfield  {journal} {\bibinfo  {journal} {J. Chem. Phys.}\
  }\textbf {\bibinfo {volume} {143}},\ \bibinfo {pages} {184502} (\bibinfo
  {year} {2015})}\BibitemShut {NoStop}%
\bibitem [{\citenamefont {Hu}\ \emph {et~al.}(2019)\citenamefont {Hu},
  \citenamefont {Schroers}, \citenamefont {Shattuck},\ and\ \citenamefont
  {O'Hern}}]{gfa_2019}%
  \BibitemOpen
  \bibfield  {author} {\bibinfo {author} {\bibfnamefont {Y.-C.}\ \bibnamefont
  {Hu}}, \bibinfo {author} {\bibfnamefont {J.}~\bibnamefont {Schroers}},
  \bibinfo {author} {\bibfnamefont {M.~D.}\ \bibnamefont {Shattuck}},\ and\
  \bibinfo {author} {\bibfnamefont {C.~S.}\ \bibnamefont {O'Hern}},\ }\bibfield
   {title} {\bibinfo {title} {Tuning the glass-forming ability of metallic
  glasses through energetic frustration},\ }\href@noop {} {\bibfield  {journal}
  {\bibinfo  {journal} {Phys. Rev. Mater.}\ }\textbf {\bibinfo {volume} {3}},\
  \bibinfo {pages} {085602} (\bibinfo {year} {2019})}\BibitemShut {NoStop}%
\bibitem [{\citenamefont {Cowley}(1960)}]{chemorder1960}%
  \BibitemOpen
  \bibfield  {author} {\bibinfo {author} {\bibfnamefont {J.~M.}\ \bibnamefont
  {Cowley}},\ }\bibfield  {title} {\bibinfo {title} {Short- and long-range
  order parameters in disordered solid solutions},\ }\href@noop {} {\bibfield
  {journal} {\bibinfo  {journal} {Phys. Rev.}\ }\textbf {\bibinfo {volume}
  {120}},\ \bibinfo {pages} {1648} (\bibinfo {year} {1960})}\BibitemShut
  {NoStop}%
\bibitem [{\citenamefont {Hu}\ and\ \citenamefont
  {Tanaka}(2020)}]{hu_physical_2020}%
  \BibitemOpen
  \bibfield  {author} {\bibinfo {author} {\bibfnamefont {Y.-C.}\ \bibnamefont
  {Hu}}\ and\ \bibinfo {author} {\bibfnamefont {H.}~\bibnamefont {Tanaka}},\
  }\bibfield  {title} {\bibinfo {title} {Physical origin of glass formation
  from multicomponent systems},\ }\href@noop {} {\bibfield  {journal} {\bibinfo
   {journal} {Sci. Adv.}\ }\textbf {\bibinfo {volume} {6}},\ \bibinfo {pages}
  {eabd2928} (\bibinfo {year} {2020})}\BibitemShut {NoStop}%
\bibitem [{\citenamefont {Steinhardt}\ \emph {et~al.}(1983)\citenamefont
  {Steinhardt}, \citenamefont {Nelson},\ and\ \citenamefont
  {Ronchetti}}]{steinhardt_bond-orientational_1983}%
  \BibitemOpen
  \bibfield  {author} {\bibinfo {author} {\bibfnamefont {P.~J.}\ \bibnamefont
  {Steinhardt}}, \bibinfo {author} {\bibfnamefont {D.~R.}\ \bibnamefont
  {Nelson}},\ and\ \bibinfo {author} {\bibfnamefont {M.}~\bibnamefont
  {Ronchetti}},\ }\bibfield  {title} {\bibinfo {title} {Bond-orientational
  order in liquids and glasses},\ }\href@noop {} {\bibfield  {journal}
  {\bibinfo  {journal} {Phys. Rev. B}\ }\textbf {\bibinfo {volume} {28}},\
  \bibinfo {pages} {784} (\bibinfo {year} {1983})}\BibitemShut {NoStop}%
\bibitem [{\citenamefont {Rycroft}\ \emph {et~al.}(2006)\citenamefont
  {Rycroft}, \citenamefont {Grest}, \citenamefont {Landry},\ and\ \citenamefont
  {Bazant}}]{rycroft_analysis_2006}%
  \BibitemOpen
  \bibfield  {author} {\bibinfo {author} {\bibfnamefont {C.~H.}\ \bibnamefont
  {Rycroft}}, \bibinfo {author} {\bibfnamefont {G.~S.}\ \bibnamefont {Grest}},
  \bibinfo {author} {\bibfnamefont {J.~W.}\ \bibnamefont {Landry}},\ and\
  \bibinfo {author} {\bibfnamefont {M.~Z.}\ \bibnamefont {Bazant}},\ }\bibfield
   {title} {\bibinfo {title} {Analysis of granular flow in a pebble-bed nuclear
  reactor},\ }\href@noop {} {\bibfield  {journal} {\bibinfo  {journal} {Phys.
  Rev. E}\ }\textbf {\bibinfo {volume} {74}},\ \bibinfo {pages} {021306}
  (\bibinfo {year} {2006})}\BibitemShut {NoStop}%
\bibitem [{\citenamefont {Rein~ten Wolde}\ \emph {et~al.}(1996)\citenamefont
  {Rein~ten Wolde}, \citenamefont {Ruiz-Montero},\ and\ \citenamefont
  {Frenkel}}]{rein_ten_wolde_numerical_1996}%
  \BibitemOpen
  \bibfield  {author} {\bibinfo {author} {\bibfnamefont {P.}~\bibnamefont
  {Rein~ten Wolde}}, \bibinfo {author} {\bibfnamefont {M.~J.}\ \bibnamefont
  {Ruiz-Montero}},\ and\ \bibinfo {author} {\bibfnamefont {D.}~\bibnamefont
  {Frenkel}},\ }\bibfield  {title} {\bibinfo {title} {Numerical calculation of
  the rate of crystal nucleation in a {Lennard}-{Jones} system at moderate
  undercooling},\ }\href@noop {} {\bibfield  {journal} {\bibinfo  {journal} {J.
  Chem. Phys.}\ }\textbf {\bibinfo {volume} {104}},\ \bibinfo {pages} {9932}
  (\bibinfo {year} {1996})}\BibitemShut {NoStop}%
\bibitem [{\citenamefont {Russo}\ and\ \citenamefont
  {Tanaka}(2012)}]{russo_microscopic_2012}%
  \BibitemOpen
  \bibfield  {author} {\bibinfo {author} {\bibfnamefont {J.}~\bibnamefont
  {Russo}}\ and\ \bibinfo {author} {\bibfnamefont {H.}~\bibnamefont {Tanaka}},\
  }\bibfield  {title} {\bibinfo {title} {The microscopic pathway to
  crystallization in supercooled liquids},\ }\href@noop {} {\bibfield
  {journal} {\bibinfo  {journal} {Sci. Rep.}\ }\textbf {\bibinfo {volume}
  {2}},\ \bibinfo {pages} {505} (\bibinfo {year} {2012})}\BibitemShut {NoStop}%
\bibitem [{\citenamefont {Sastry}\ \emph {et~al.}(1998)\citenamefont {Sastry},
  \citenamefont {Debenedetti},\ and\ \citenamefont
  {Stillinger}}]{sastry_signatures_1998}%
  \BibitemOpen
  \bibfield  {author} {\bibinfo {author} {\bibfnamefont {S.}~\bibnamefont
  {Sastry}}, \bibinfo {author} {\bibfnamefont {P.~G.}\ \bibnamefont
  {Debenedetti}},\ and\ \bibinfo {author} {\bibfnamefont {F.~H.}\ \bibnamefont
  {Stillinger}},\ }\bibfield  {title} {\bibinfo {title} {Signatures of distinct
  dynamical regimes in the energy landscape of a glass-forming liquid},\
  }\href@noop {} {\bibfield  {journal} {\bibinfo  {journal} {Nature}\ }\textbf
  {\bibinfo {volume} {393}},\ \bibinfo {pages} {554} (\bibinfo {year}
  {1998})}\BibitemShut {NoStop}%
\bibitem [{\citenamefont {Debenedetti}\ and\ \citenamefont
  {Stillinger}(2001)}]{debenedetti_supercooled_2001}%
  \BibitemOpen
  \bibfield  {author} {\bibinfo {author} {\bibfnamefont {P.~G.}\ \bibnamefont
  {Debenedetti}}\ and\ \bibinfo {author} {\bibfnamefont {F.~H.}\ \bibnamefont
  {Stillinger}},\ }\bibfield  {title} {\bibinfo {title} {Supercooled liquids
  and the glass transition},\ }\href@noop {} {\bibfield  {journal} {\bibinfo
  {journal} {Nature}\ }\textbf {\bibinfo {volume} {410}},\ \bibinfo {pages}
  {259} (\bibinfo {year} {2001})}\BibitemShut {NoStop}%
\end{thebibliography}
%

\balance
\end{document}